\documentclass[longauth]{aaEC}

\usepackage{graphicx}
\usepackage{natbib}
\usepackage{scalerel}
\usepackage{xspace}
\usepackage{ulem}

\usepackage[table]{xcolor}

\usepackage{txfonts}

\usepackage[pdfencoding=auto,psdextra]{hyperref}
\hypersetup{
    colorlinks=true,
    linkcolor=blue,
    filecolor=magenta,      
    urlcolor=blue,
    citecolor=blue
}
\makeatletter
\renewcommand*\aa@pageof{, page \thepage{} of \pageref*{LastPage}}
\makeatother

\usepackage[utf8]{inputenc}

\usepackage[switch, modulo]{lineno}

\usepackage{euclid}

\newcommand{\fha}{\ensuremath{f_{{\rm H}\alpha}}\xspace}

\newcommand{\hmpc}{\ensuremath{h^{-1}\,\text{Mpc}}\xspace}
\newcommand{\kmpc}{\ensuremath{h\,\text{Mpc}^{-1}}\xspace}
\newcommand{\dens}{\ensuremath{h^{3}\,\mathrm{Mpc}^{-3}}\xspace}
\newcommand{\cgpc}{\ensuremath{h^{-3}\,\mathrm{Gpc}^3}\xspace}
\newcommand{\hgpc}{\ensuremath{h^{-1}\,\text{Gpc}}\xspace}
\newcommand{\hmsun}{\ensuremath{h^{-1}\,M_\odot}\xspace}

\newcommand{\flux}{erg s$^{-1}$ cm$^{-2}$\xspace}
\newcommand{\pin}{\texttt{Pinocchio}\xspace}
\newcommand{\tdeg}{\ensuremath{30^{\circ}}\xspace}
\newcommand{\sqdeg}{\ensuremath{{\rm deg}^2}\xspace}

\begin{document}
\title{\Euclid preparation}
\subtitle{LXXVI. Simulating thousands of \Euclid spectroscopic skies}

\newcommand{\orcid}[1]{} 	   
\author{Euclid Collaboration: P.~Monaco\orcid{0000-0003-2083-7564}\thanks{\email{pierluigi.monaco@inaf.it}}\inst{\ref{aff1},\ref{aff2},\ref{aff3},\ref{aff4}}
\and G.~Parimbelli\orcid{0000-0002-2539-2472}\inst{\ref{aff5},\ref{aff6},\ref{aff7}}
\and M.~Y.~Elkhashab\orcid{0000-0001-9306-2603}\inst{\ref{aff2},\ref{aff3},\ref{aff1},\ref{aff8}}
\and J.~Salvalaggio\orcid{0000-0002-1431-5607}\inst{\ref{aff2},\ref{aff8},\ref{aff1},\ref{aff3}}
\and T.~Castro\orcid{0000-0002-6292-3228}\inst{\ref{aff2},\ref{aff3},\ref{aff8},\ref{aff4}}
\and M.~D.~Lepinzan\orcid{0000-0003-1287-9801}\inst{\ref{aff1},\ref{aff2}}
\and E.~Sarpa\orcid{0000-0002-1256-655X}\inst{\ref{aff7},\ref{aff4},\ref{aff3}}
\and E.~Sefusatti\orcid{0000-0003-0473-1567}\inst{\ref{aff2},\ref{aff8},\ref{aff3}}
\and L.~Stanco\orcid{0000-0002-9706-5104}\inst{\ref{aff9}}
\and L.~Tornatore\orcid{0000-0003-1751-0130}\inst{\ref{aff2}}
\and G.~E.~Addison\orcid{0000-0002-2147-2248}\inst{\ref{aff10}}
\and S.~Bruton\orcid{0000-0002-6503-5218}\inst{\ref{aff11}}
\and C.~Carbone\orcid{0000-0003-0125-3563}\inst{\ref{aff12}}
\and F.~J.~Castander\orcid{0000-0001-7316-4573}\inst{\ref{aff5},\ref{aff13}}
\and J.~Carretero\orcid{0000-0002-3130-0204}\inst{\ref{aff14},\ref{aff15}}
\and S.~de~la~Torre\inst{\ref{aff16}}
\and P.~Fosalba\orcid{0000-0002-1510-5214}\inst{\ref{aff13},\ref{aff5}}
\and G.~Lavaux\orcid{0000-0003-0143-8891}\inst{\ref{aff17}}
\and S.~Lee\orcid{0000-0002-8289-740X}\inst{\ref{aff18}}
\and K.~Markovic\orcid{0000-0001-6764-073X}\inst{\ref{aff18}}
\and K.~S.~McCarthy\orcid{0000-0001-6857-018X}\inst{\ref{aff18},\ref{aff19}}
\and F.~Passalacqua\orcid{0000-0002-8606-4093}\inst{\ref{aff20},\ref{aff9}}
\and W.~J.~Percival\orcid{0000-0002-0644-5727}\inst{\ref{aff21},\ref{aff22},\ref{aff23}}
\and I.~Risso\orcid{0000-0003-2525-7761}\inst{\ref{aff24},\ref{aff25}}
\and C.~Scarlata\orcid{0000-0002-9136-8876}\inst{\ref{aff26}}
\and P.~Tallada-Cresp\'{i}\orcid{0000-0002-1336-8328}\inst{\ref{aff14},\ref{aff15}}
\and M.~Viel\orcid{0000-0002-2642-5707}\inst{\ref{aff8},\ref{aff2},\ref{aff7},\ref{aff3},\ref{aff4}}
\and Y.~Wang\orcid{0000-0002-4749-2984}\inst{\ref{aff27}}
\and B.~Altieri\orcid{0000-0003-3936-0284}\inst{\ref{aff28}}
\and S.~Andreon\orcid{0000-0002-2041-8784}\inst{\ref{aff24}}
\and N.~Auricchio\orcid{0000-0003-4444-8651}\inst{\ref{aff29}}
\and C.~Baccigalupi\orcid{0000-0002-8211-1630}\inst{\ref{aff8},\ref{aff2},\ref{aff3},\ref{aff7}}
\and M.~Baldi\orcid{0000-0003-4145-1943}\inst{\ref{aff30},\ref{aff29},\ref{aff31}}
\and S.~Bardelli\orcid{0000-0002-8900-0298}\inst{\ref{aff29}}
\and P.~Battaglia\orcid{0000-0002-7337-5909}\inst{\ref{aff29}}
\and F.~Bernardeau\orcid{0009-0007-3015-2581}\inst{\ref{aff32},\ref{aff17}}
\and A.~Biviano\orcid{0000-0002-0857-0732}\inst{\ref{aff2},\ref{aff8}}
\and E.~Branchini\orcid{0000-0002-0808-6908}\inst{\ref{aff33},\ref{aff25},\ref{aff24}}
\and M.~Brescia\orcid{0000-0001-9506-5680}\inst{\ref{aff34},\ref{aff35}}
\and J.~Brinchmann\orcid{0000-0003-4359-8797}\inst{\ref{aff36},\ref{aff37},\ref{aff38}}
\and S.~Camera\orcid{0000-0003-3399-3574}\inst{\ref{aff39},\ref{aff40},\ref{aff41}}
\and G.~Ca\~nas-Herrera\orcid{0000-0003-2796-2149}\inst{\ref{aff42},\ref{aff43},\ref{aff44}}
\and V.~Capobianco\orcid{0000-0002-3309-7692}\inst{\ref{aff41}}
\and V.~F.~Cardone\inst{\ref{aff45},\ref{aff46}}
\and S.~Casas\orcid{0000-0002-4751-5138}\inst{\ref{aff47}}
\and M.~Castellano\orcid{0000-0001-9875-8263}\inst{\ref{aff45}}
\and G.~Castignani\orcid{0000-0001-6831-0687}\inst{\ref{aff29}}
\and S.~Cavuoti\orcid{0000-0002-3787-4196}\inst{\ref{aff35},\ref{aff48}}
\and A.~Cimatti\inst{\ref{aff49}}
\and C.~Colodro-Conde\inst{\ref{aff50}}
\and G.~Congedo\orcid{0000-0003-2508-0046}\inst{\ref{aff51}}
\and C.~J.~Conselice\orcid{0000-0003-1949-7638}\inst{\ref{aff52}}
\and L.~Conversi\orcid{0000-0002-6710-8476}\inst{\ref{aff53},\ref{aff28}}
\and Y.~Copin\orcid{0000-0002-5317-7518}\inst{\ref{aff54}}
\and F.~Courbin\orcid{0000-0003-0758-6510}\inst{\ref{aff55},\ref{aff56}}
\and H.~M.~Courtois\orcid{0000-0003-0509-1776}\inst{\ref{aff57}}
\and A.~Da~Silva\orcid{0000-0002-6385-1609}\inst{\ref{aff58},\ref{aff59}}
\and H.~Degaudenzi\orcid{0000-0002-5887-6799}\inst{\ref{aff60}}
\and G.~De~Lucia\orcid{0000-0002-6220-9104}\inst{\ref{aff2}}
\and A.~M.~Di~Giorgio\orcid{0000-0002-4767-2360}\inst{\ref{aff61}}
\and F.~Dubath\orcid{0000-0002-6533-2810}\inst{\ref{aff60}}
\and F.~Ducret\inst{\ref{aff16}}
\and C.~A.~J.~Duncan\orcid{0009-0003-3573-0791}\inst{\ref{aff51},\ref{aff52}}
\and X.~Dupac\inst{\ref{aff28}}
\and S.~Dusini\orcid{0000-0002-1128-0664}\inst{\ref{aff9}}
\and A.~Ealet\orcid{0000-0003-3070-014X}\inst{\ref{aff54}}
\and S.~Escoffier\orcid{0000-0002-2847-7498}\inst{\ref{aff62}}
\and M.~Farina\orcid{0000-0002-3089-7846}\inst{\ref{aff61}}
\and R.~Farinelli\inst{\ref{aff29}}
\and S.~Farrens\orcid{0000-0002-9594-9387}\inst{\ref{aff63}}
\and S.~Ferriol\inst{\ref{aff54}}
\and F.~Finelli\orcid{0000-0002-6694-3269}\inst{\ref{aff29},\ref{aff64}}
\and N.~Fourmanoit\orcid{0009-0005-6816-6925}\inst{\ref{aff62}}
\and M.~Frailis\orcid{0000-0002-7400-2135}\inst{\ref{aff2}}
\and E.~Franceschi\orcid{0000-0002-0585-6591}\inst{\ref{aff29}}
\and M.~Fumana\orcid{0000-0001-6787-5950}\inst{\ref{aff12}}
\and S.~Galeotta\orcid{0000-0002-3748-5115}\inst{\ref{aff2}}
\and K.~George\orcid{0000-0002-1734-8455}\inst{\ref{aff65}}
\and B.~Gillis\orcid{0000-0002-4478-1270}\inst{\ref{aff51}}
\and C.~Giocoli\orcid{0000-0002-9590-7961}\inst{\ref{aff29},\ref{aff31}}
\and J.~Gracia-Carpio\inst{\ref{aff66}}
\and A.~Grazian\orcid{0000-0002-5688-0663}\inst{\ref{aff67}}
\and F.~Grupp\inst{\ref{aff66},\ref{aff68}}
\and L.~Guzzo\orcid{0000-0001-8264-5192}\inst{\ref{aff69},\ref{aff24},\ref{aff70}}
\and S.~V.~H.~Haugan\orcid{0000-0001-9648-7260}\inst{\ref{aff71}}
\and W.~Holmes\inst{\ref{aff18}}
\and F.~Hormuth\inst{\ref{aff72}}
\and A.~Hornstrup\orcid{0000-0002-3363-0936}\inst{\ref{aff73},\ref{aff74}}
\and K.~Jahnke\orcid{0000-0003-3804-2137}\inst{\ref{aff75}}
\and M.~Jhabvala\inst{\ref{aff76}}
\and B.~Joachimi\orcid{0000-0001-7494-1303}\inst{\ref{aff77}}
\and E.~Keih\"anen\orcid{0000-0003-1804-7715}\inst{\ref{aff78}}
\and S.~Kermiche\orcid{0000-0002-0302-5735}\inst{\ref{aff62}}
\and B.~Kubik\orcid{0009-0006-5823-4880}\inst{\ref{aff54}}
\and M.~K\"ummel\orcid{0000-0003-2791-2117}\inst{\ref{aff68}}
\and M.~Kunz\orcid{0000-0002-3052-7394}\inst{\ref{aff79}}
\and H.~Kurki-Suonio\orcid{0000-0002-4618-3063}\inst{\ref{aff80},\ref{aff81}}
\and A.~M.~C.~Le~Brun\orcid{0000-0002-0936-4594}\inst{\ref{aff82}}
\and S.~Ligori\orcid{0000-0003-4172-4606}\inst{\ref{aff41}}
\and P.~B.~Lilje\orcid{0000-0003-4324-7794}\inst{\ref{aff71}}
\and V.~Lindholm\orcid{0000-0003-2317-5471}\inst{\ref{aff80},\ref{aff81}}
\and I.~Lloro\orcid{0000-0001-5966-1434}\inst{\ref{aff83}}
\and D.~Maino\inst{\ref{aff69},\ref{aff12},\ref{aff70}}
\and E.~Maiorano\orcid{0000-0003-2593-4355}\inst{\ref{aff29}}
\and O.~Mansutti\orcid{0000-0001-5758-4658}\inst{\ref{aff2}}
\and O.~Marggraf\orcid{0000-0001-7242-3852}\inst{\ref{aff84}}
\and M.~Martinelli\orcid{0000-0002-6943-7732}\inst{\ref{aff45},\ref{aff46}}
\and N.~Martinet\orcid{0000-0003-2786-7790}\inst{\ref{aff16}}
\and F.~Marulli\orcid{0000-0002-8850-0303}\inst{\ref{aff85},\ref{aff29},\ref{aff31}}
\and R.~Massey\orcid{0000-0002-6085-3780}\inst{\ref{aff86}}
\and E.~Medinaceli\orcid{0000-0002-4040-7783}\inst{\ref{aff29}}
\and S.~Mei\orcid{0000-0002-2849-559X}\inst{\ref{aff87},\ref{aff88}}
\and M.~Melchior\inst{\ref{aff89}}
\and Y.~Mellier\inst{\ref{aff90},\ref{aff17}}
\and M.~Meneghetti\orcid{0000-0003-1225-7084}\inst{\ref{aff29},\ref{aff31}}
\and E.~Merlin\orcid{0000-0001-6870-8900}\inst{\ref{aff45}}
\and G.~Meylan\inst{\ref{aff91}}
\and A.~Mora\orcid{0000-0002-1922-8529}\inst{\ref{aff92}}
\and M.~Moresco\orcid{0000-0002-7616-7136}\inst{\ref{aff85},\ref{aff29}}
\and L.~Moscardini\orcid{0000-0002-3473-6716}\inst{\ref{aff85},\ref{aff29},\ref{aff31}}
\and E.~Munari\orcid{0000-0002-1751-5946}\inst{\ref{aff2},\ref{aff8}}
\and R.~Nakajima\orcid{0009-0009-1213-7040}\inst{\ref{aff84}}
\and C.~Neissner\orcid{0000-0001-8524-4968}\inst{\ref{aff93},\ref{aff15}}
\and S.-M.~Niemi\orcid{0009-0005-0247-0086}\inst{\ref{aff42}}
\and C.~Padilla\orcid{0000-0001-7951-0166}\inst{\ref{aff93}}
\and S.~Paltani\orcid{0000-0002-8108-9179}\inst{\ref{aff60}}
\and F.~Pasian\orcid{0000-0002-4869-3227}\inst{\ref{aff2}}
\and K.~Pedersen\inst{\ref{aff94}}
\and V.~Pettorino\inst{\ref{aff42}}
\and S.~Pires\orcid{0000-0002-0249-2104}\inst{\ref{aff63}}
\and G.~Polenta\orcid{0000-0003-4067-9196}\inst{\ref{aff95}}
\and M.~Poncet\inst{\ref{aff96}}
\and L.~A.~Popa\inst{\ref{aff97}}
\and L.~Pozzetti\orcid{0000-0001-7085-0412}\inst{\ref{aff29}}
\and F.~Raison\orcid{0000-0002-7819-6918}\inst{\ref{aff66}}
\and A.~Renzi\orcid{0000-0001-9856-1970}\inst{\ref{aff20},\ref{aff9}}
\and J.~Rhodes\orcid{0000-0002-4485-8549}\inst{\ref{aff18}}
\and G.~Riccio\inst{\ref{aff35}}
\and F.~Rizzo\orcid{0000-0002-9407-585X}\inst{\ref{aff2}}
\and E.~Romelli\orcid{0000-0003-3069-9222}\inst{\ref{aff2}}
\and M.~Roncarelli\orcid{0000-0001-9587-7822}\inst{\ref{aff29}}
\and R.~Saglia\orcid{0000-0003-0378-7032}\inst{\ref{aff68},\ref{aff66}}
\and Z.~Sakr\orcid{0000-0002-4823-3757}\inst{\ref{aff98},\ref{aff99},\ref{aff100}}
\and A.~G.~S\'anchez\orcid{0000-0003-1198-831X}\inst{\ref{aff66}}
\and D.~Sapone\orcid{0000-0001-7089-4503}\inst{\ref{aff101}}
\and B.~Sartoris\orcid{0000-0003-1337-5269}\inst{\ref{aff68},\ref{aff2}}
\and P.~Schneider\orcid{0000-0001-8561-2679}\inst{\ref{aff84}}
\and T.~Schrabback\orcid{0000-0002-6987-7834}\inst{\ref{aff102},\ref{aff84}}
\and M.~Scodeggio\inst{\ref{aff12}}
\and A.~Secroun\orcid{0000-0003-0505-3710}\inst{\ref{aff62}}
\and G.~Seidel\orcid{0000-0003-2907-353X}\inst{\ref{aff75}}
\and M.~Seiffert\orcid{0000-0002-7536-9393}\inst{\ref{aff18}}
\and S.~Serrano\orcid{0000-0002-0211-2861}\inst{\ref{aff13},\ref{aff103},\ref{aff5}}
\and P.~Simon\inst{\ref{aff84}}
\and C.~Sirignano\orcid{0000-0002-0995-7146}\inst{\ref{aff20},\ref{aff9}}
\and G.~Sirri\orcid{0000-0003-2626-2853}\inst{\ref{aff31}}
\and J.~Steinwagner\orcid{0000-0001-7443-1047}\inst{\ref{aff66}}
\and D.~Tavagnacco\orcid{0000-0001-7475-9894}\inst{\ref{aff2}}
\and A.~N.~Taylor\inst{\ref{aff51}}
\and I.~Tereno\orcid{0000-0002-4537-6218}\inst{\ref{aff58},\ref{aff104}}
\and N.~Tessore\orcid{0000-0002-9696-7931}\inst{\ref{aff77}}
\and S.~Toft\orcid{0000-0003-3631-7176}\inst{\ref{aff105},\ref{aff106}}
\and R.~Toledo-Moreo\orcid{0000-0002-2997-4859}\inst{\ref{aff107}}
\and F.~Torradeflot\orcid{0000-0003-1160-1517}\inst{\ref{aff15},\ref{aff14}}
\and I.~Tutusaus\orcid{0000-0002-3199-0399}\inst{\ref{aff99}}
\and L.~Valenziano\orcid{0000-0002-1170-0104}\inst{\ref{aff29},\ref{aff64}}
\and J.~Valiviita\orcid{0000-0001-6225-3693}\inst{\ref{aff80},\ref{aff81}}
\and T.~Vassallo\orcid{0000-0001-6512-6358}\inst{\ref{aff68},\ref{aff2}}
\and G.~Verdoes~Kleijn\orcid{0000-0001-5803-2580}\inst{\ref{aff108}}
\and A.~Veropalumbo\orcid{0000-0003-2387-1194}\inst{\ref{aff24},\ref{aff25},\ref{aff33}}
\and J.~Weller\orcid{0000-0002-8282-2010}\inst{\ref{aff68},\ref{aff66}}
\and G.~Zamorani\orcid{0000-0002-2318-301X}\inst{\ref{aff29}}
\and E.~Zucca\orcid{0000-0002-5845-8132}\inst{\ref{aff29}}
\and V.~Allevato\orcid{0000-0001-7232-5152}\inst{\ref{aff35}}
\and M.~Ballardini\orcid{0000-0003-4481-3559}\inst{\ref{aff109},\ref{aff110},\ref{aff29}}
\and C.~Burigana\orcid{0000-0002-3005-5796}\inst{\ref{aff111},\ref{aff64}}
\and R.~Cabanac\orcid{0000-0001-6679-2600}\inst{\ref{aff99}}
\and M.~Calabrese\orcid{0000-0002-2637-2422}\inst{\ref{aff112},\ref{aff12}}
\and A.~Cappi\inst{\ref{aff29},\ref{aff113}}
\and D.~Di~Ferdinando\inst{\ref{aff31}}
\and J.~A.~Escartin~Vigo\inst{\ref{aff66}}
\and G.~Fabbian\orcid{0000-0002-3255-4695}\inst{\ref{aff114},\ref{aff115}}
\and L.~Gabarra\orcid{0000-0002-8486-8856}\inst{\ref{aff116}}
\and J.~Mart\'{i}n-Fleitas\orcid{0000-0002-8594-569X}\inst{\ref{aff117}}
\and S.~Matthew\orcid{0000-0001-8448-1697}\inst{\ref{aff51}}
\and N.~Mauri\orcid{0000-0001-8196-1548}\inst{\ref{aff49},\ref{aff31}}
\and R.~B.~Metcalf\orcid{0000-0003-3167-2574}\inst{\ref{aff85},\ref{aff29}}
\and A.~Pezzotta\orcid{0000-0003-0726-2268}\inst{\ref{aff24}}
\and M.~P\"ontinen\orcid{0000-0001-5442-2530}\inst{\ref{aff80}}
\and C.~Porciani\orcid{0000-0002-7797-2508}\inst{\ref{aff84}}
\and V.~Scottez\orcid{0009-0008-3864-940X}\inst{\ref{aff90},\ref{aff118}}
\and M.~Sereno\orcid{0000-0003-0302-0325}\inst{\ref{aff29},\ref{aff31}}
\and M.~Tenti\orcid{0000-0002-4254-5901}\inst{\ref{aff31}}
\and M.~Wiesmann\orcid{0009-0000-8199-5860}\inst{\ref{aff71}}
\and Y.~Akrami\orcid{0000-0002-2407-7956}\inst{\ref{aff119},\ref{aff120}}
\and S.~Alvi\orcid{0000-0001-5779-8568}\inst{\ref{aff109}}
\and I.~T.~Andika\orcid{0000-0001-6102-9526}\inst{\ref{aff121},\ref{aff122}}
\and S.~Anselmi\orcid{0000-0002-3579-9583}\inst{\ref{aff9},\ref{aff20},\ref{aff123}}
\and M.~Archidiacono\orcid{0000-0003-4952-9012}\inst{\ref{aff69},\ref{aff70}}
\and F.~Atrio-Barandela\orcid{0000-0002-2130-2513}\inst{\ref{aff124}}
\and S.~Avila\orcid{0000-0001-5043-3662}\inst{\ref{aff14}}
\and A.~Balaguera-Antolinez\orcid{0000-0001-5028-3035}\inst{\ref{aff50}}
\and P.~Bergamini\orcid{0000-0003-1383-9414}\inst{\ref{aff69},\ref{aff29}}
\and D.~Bertacca\orcid{0000-0002-2490-7139}\inst{\ref{aff20},\ref{aff67},\ref{aff9}}
\and M.~Bethermin\orcid{0000-0002-3915-2015}\inst{\ref{aff125}}
\and A.~Blanchard\orcid{0000-0001-8555-9003}\inst{\ref{aff99}}
\and L.~Blot\orcid{0000-0002-9622-7167}\inst{\ref{aff126},\ref{aff82}}
\and S.~Borgani\orcid{0000-0001-6151-6439}\inst{\ref{aff1},\ref{aff8},\ref{aff2},\ref{aff3},\ref{aff4}}
\and M.~L.~Brown\orcid{0000-0002-0370-8077}\inst{\ref{aff52}}
\and A.~Calabro\orcid{0000-0003-2536-1614}\inst{\ref{aff45}}
\and B.~Camacho~Quevedo\orcid{0000-0002-8789-4232}\inst{\ref{aff8},\ref{aff7},\ref{aff2}}
\and F.~Caro\inst{\ref{aff45}}
\and C.~S.~Carvalho\inst{\ref{aff104}}
\and F.~Cogato\orcid{0000-0003-4632-6113}\inst{\ref{aff85},\ref{aff29}}
\and S.~Conseil\orcid{0000-0002-3657-4191}\inst{\ref{aff54}}
\and S.~Contarini\orcid{0000-0002-9843-723X}\inst{\ref{aff66}}
\and A.~R.~Cooray\orcid{0000-0002-3892-0190}\inst{\ref{aff127}}
\and O.~Cucciati\orcid{0000-0002-9336-7551}\inst{\ref{aff29}}
\and S.~Davini\orcid{0000-0003-3269-1718}\inst{\ref{aff25}}
\and G.~Desprez\orcid{0000-0001-8325-1742}\inst{\ref{aff108}}
\and A.~D\'iaz-S\'anchez\orcid{0000-0003-0748-4768}\inst{\ref{aff128}}
\and J.~J.~Diaz\orcid{0000-0003-2101-1078}\inst{\ref{aff50}}
\and S.~Di~Domizio\orcid{0000-0003-2863-5895}\inst{\ref{aff33},\ref{aff25}}
\and J.~M.~Diego\orcid{0000-0001-9065-3926}\inst{\ref{aff129}}
\and A.~Enia\orcid{0000-0002-0200-2857}\inst{\ref{aff30},\ref{aff29}}
\and Y.~Fang\inst{\ref{aff68}}
\and A.~G.~Ferrari\orcid{0009-0005-5266-4110}\inst{\ref{aff31}}
\and A.~Finoguenov\orcid{0000-0002-4606-5403}\inst{\ref{aff80}}
\and F.~Fontanot\orcid{0000-0003-4744-0188}\inst{\ref{aff2},\ref{aff8}}
\and A.~Franco\orcid{0000-0002-4761-366X}\inst{\ref{aff130},\ref{aff131},\ref{aff132}}
\and K.~Ganga\orcid{0000-0001-8159-8208}\inst{\ref{aff87}}
\and J.~Garc\'ia-Bellido\orcid{0000-0002-9370-8360}\inst{\ref{aff119}}
\and T.~Gasparetto\orcid{0000-0002-7913-4866}\inst{\ref{aff45}}
\and V.~Gautard\inst{\ref{aff133}}
\and E.~Gaztanaga\orcid{0000-0001-9632-0815}\inst{\ref{aff5},\ref{aff13},\ref{aff134}}
\and F.~Giacomini\orcid{0000-0002-3129-2814}\inst{\ref{aff31}}
\and F.~Gianotti\orcid{0000-0003-4666-119X}\inst{\ref{aff29}}
\and G.~Gozaliasl\orcid{0000-0002-0236-919X}\inst{\ref{aff135},\ref{aff80}}
\and M.~Guidi\orcid{0000-0001-9408-1101}\inst{\ref{aff30},\ref{aff29}}
\and C.~M.~Gutierrez\orcid{0000-0001-7854-783X}\inst{\ref{aff136}}
\and A.~Hall\orcid{0000-0002-3139-8651}\inst{\ref{aff51}}
\and S.~Hemmati\orcid{0000-0003-2226-5395}\inst{\ref{aff137}}
\and C.~Hern\'andez-Monteagudo\orcid{0000-0001-5471-9166}\inst{\ref{aff138},\ref{aff50}}
\and H.~Hildebrandt\orcid{0000-0002-9814-3338}\inst{\ref{aff139}}
\and J.~Hjorth\orcid{0000-0002-4571-2306}\inst{\ref{aff94}}
\and S.~Joudaki\orcid{0000-0001-8820-673X}\inst{\ref{aff14}}
\and J.~J.~E.~Kajava\orcid{0000-0002-3010-8333}\inst{\ref{aff140},\ref{aff141}}
\and Y.~Kang\orcid{0009-0000-8588-7250}\inst{\ref{aff60}}
\and V.~Kansal\orcid{0000-0002-4008-6078}\inst{\ref{aff142},\ref{aff143}}
\and D.~Karagiannis\orcid{0000-0002-4927-0816}\inst{\ref{aff109},\ref{aff144}}
\and K.~Kiiveri\inst{\ref{aff78}}
\and C.~C.~Kirkpatrick\inst{\ref{aff78}}
\and S.~Kruk\orcid{0000-0001-8010-8879}\inst{\ref{aff28}}
\and V.~Le~Brun\orcid{0000-0002-5027-1939}\inst{\ref{aff16}}
\and J.~Le~Graet\orcid{0000-0001-6523-7971}\inst{\ref{aff62}}
\and L.~Legrand\orcid{0000-0003-0610-5252}\inst{\ref{aff145},\ref{aff146}}
\and M.~Lembo\orcid{0000-0002-5271-5070}\inst{\ref{aff17}}
\and F.~Lepori\orcid{0009-0000-5061-7138}\inst{\ref{aff147}}
\and G.~Leroy\orcid{0009-0004-2523-4425}\inst{\ref{aff148},\ref{aff86}}
\and G.~F.~Lesci\orcid{0000-0002-4607-2830}\inst{\ref{aff85},\ref{aff29}}
\and J.~Lesgourgues\orcid{0000-0001-7627-353X}\inst{\ref{aff47}}
\and L.~Leuzzi\orcid{0009-0006-4479-7017}\inst{\ref{aff29}}
\and T.~I.~Liaudat\orcid{0000-0002-9104-314X}\inst{\ref{aff149}}
\and J.~Macias-Perez\orcid{0000-0002-5385-2763}\inst{\ref{aff150}}
\and G.~Maggio\orcid{0000-0003-4020-4836}\inst{\ref{aff2}}
\and M.~Magliocchetti\orcid{0000-0001-9158-4838}\inst{\ref{aff61}}
\and C.~Mancini\orcid{0000-0002-4297-0561}\inst{\ref{aff12}}
\and F.~Mannucci\orcid{0000-0002-4803-2381}\inst{\ref{aff151}}
\and R.~Maoli\orcid{0000-0002-6065-3025}\inst{\ref{aff152},\ref{aff45}}
\and C.~J.~A.~P.~Martins\orcid{0000-0002-4886-9261}\inst{\ref{aff153},\ref{aff36}}
\and L.~Maurin\orcid{0000-0002-8406-0857}\inst{\ref{aff114}}
\and M.~Miluzio\inst{\ref{aff28},\ref{aff154}}
\and A.~Montoro\orcid{0000-0003-4730-8590}\inst{\ref{aff5},\ref{aff13}}
\and C.~Moretti\orcid{0000-0003-3314-8936}\inst{\ref{aff2},\ref{aff8},\ref{aff7},\ref{aff3}}
\and G.~Morgante\inst{\ref{aff29}}
\and S.~Nadathur\orcid{0000-0001-9070-3102}\inst{\ref{aff134}}
\and K.~Naidoo\orcid{0000-0002-9182-1802}\inst{\ref{aff134}}
\and A.~Navarro-Alsina\orcid{0000-0002-3173-2592}\inst{\ref{aff84}}
\and S.~Nesseris\orcid{0000-0002-0567-0324}\inst{\ref{aff119}}
\and K.~Paterson\orcid{0000-0001-8340-3486}\inst{\ref{aff75}}
\and A.~Pisani\orcid{0000-0002-6146-4437}\inst{\ref{aff62}}
\and D.~Potter\orcid{0000-0002-0757-5195}\inst{\ref{aff147}}
\and S.~Quai\orcid{0000-0002-0449-8163}\inst{\ref{aff85},\ref{aff29}}
\and M.~Radovich\orcid{0000-0002-3585-866X}\inst{\ref{aff67}}
\and G.~Rodighiero\orcid{0000-0002-9415-2296}\inst{\ref{aff20},\ref{aff67}}
\and S.~Sacquegna\orcid{0000-0002-8433-6630}\inst{\ref{aff155},\ref{aff131},\ref{aff130}}
\and M.~Sahl\'en\orcid{0000-0003-0973-4804}\inst{\ref{aff156}}
\and D.~B.~Sanders\orcid{0000-0002-1233-9998}\inst{\ref{aff157}}
\and D.~Sciotti\orcid{0009-0008-4519-2620}\inst{\ref{aff45},\ref{aff46}}
\and E.~Sellentin\inst{\ref{aff158},\ref{aff44}}
\and L.~C.~Smith\orcid{0000-0002-3259-2771}\inst{\ref{aff159}}
\and J.~G.~Sorce\orcid{0000-0002-2307-2432}\inst{\ref{aff160},\ref{aff114}}
\and K.~Tanidis\orcid{0000-0001-9843-5130}\inst{\ref{aff116}}
\and C.~Tao\orcid{0000-0001-7961-8177}\inst{\ref{aff62}}
\and G.~Testera\inst{\ref{aff25}}
\and R.~Teyssier\orcid{0000-0001-7689-0933}\inst{\ref{aff161}}
\and S.~Tosi\orcid{0000-0002-7275-9193}\inst{\ref{aff33},\ref{aff25},\ref{aff24}}
\and A.~Troja\orcid{0000-0003-0239-4595}\inst{\ref{aff20},\ref{aff9}}
\and M.~Tucci\inst{\ref{aff60}}
\and C.~Valieri\inst{\ref{aff31}}
\and A.~Venhola\orcid{0000-0001-6071-4564}\inst{\ref{aff162}}
\and F.~Vernizzi\orcid{0000-0003-3426-2802}\inst{\ref{aff32}}
\and G.~Verza\orcid{0000-0002-1886-8348}\inst{\ref{aff163}}
\and P.~Vielzeuf\orcid{0000-0003-2035-9339}\inst{\ref{aff62}}
\and N.~A.~Walton\orcid{0000-0003-3983-8778}\inst{\ref{aff159}}}

\institute{Dipartimento di Fisica - Sezione di Astronomia, Universit\`a di Trieste, Via Tiepolo 11, 34131 Trieste, Italy\label{aff1}
\and
INAF-Osservatorio Astronomico di Trieste, Via G. B. Tiepolo 11, 34143 Trieste, Italy\label{aff2}
\and
INFN, Sezione di Trieste, Via Valerio 2, 34127 Trieste TS, Italy\label{aff3}
\and
ICSC - Centro Nazionale di Ricerca in High Performance Computing, Big Data e Quantum Computing, Via Magnanelli 2, Bologna, Italy\label{aff4}
\and
Institute of Space Sciences (ICE, CSIC), Campus UAB, Carrer de Can Magrans, s/n, 08193 Barcelona, Spain\label{aff5}
\and
Dipartimento di Fisica, Universit\`a degli studi di Genova, and INFN-Sezione di Genova, via Dodecaneso 33, 16146, Genova, Italy\label{aff6}
\and
SISSA, International School for Advanced Studies, Via Bonomea 265, 34136 Trieste TS, Italy\label{aff7}
\and
IFPU, Institute for Fundamental Physics of the Universe, via Beirut 2, 34151 Trieste, Italy\label{aff8}
\and
INFN-Padova, Via Marzolo 8, 35131 Padova, Italy\label{aff9}
\and
Johns Hopkins University 3400 North Charles Street Baltimore, MD 21218, USA\label{aff10}
\and
California Institute of Technology, 1200 E California Blvd, Pasadena, CA 91125, USA\label{aff11}
\and
INAF-IASF Milano, Via Alfonso Corti 12, 20133 Milano, Italy\label{aff12}
\and
Institut d'Estudis Espacials de Catalunya (IEEC),  Edifici RDIT, Campus UPC, 08860 Castelldefels, Barcelona, Spain\label{aff13}
\and
Centro de Investigaciones Energ\'eticas, Medioambientales y Tecnol\'ogicas (CIEMAT), Avenida Complutense 40, 28040 Madrid, Spain\label{aff14}
\and
Port d'Informaci\'{o} Cient\'{i}fica, Campus UAB, C. Albareda s/n, 08193 Bellaterra (Barcelona), Spain\label{aff15}
\and
Aix-Marseille Universit\'e, CNRS, CNES, LAM, Marseille, France\label{aff16}
\and
Institut d'Astrophysique de Paris, UMR 7095, CNRS, and Sorbonne Universit\'e, 98 bis boulevard Arago, 75014 Paris, France\label{aff17}
\and
Jet Propulsion Laboratory, California Institute of Technology, 4800 Oak Grove Drive, Pasadena, CA, 91109, USA\label{aff18}
\and
Kavli Institute for the Physics and Mathematics of the Universe (WPI), University of Tokyo, Kashiwa, Chiba 277-8583, Japan\label{aff19}
\and
Dipartimento di Fisica e Astronomia "G. Galilei", Universit\`a di Padova, Via Marzolo 8, 35131 Padova, Italy\label{aff20}
\and
Waterloo Centre for Astrophysics, University of Waterloo, Waterloo, Ontario N2L 3G1, Canada\label{aff21}
\and
Department of Physics and Astronomy, University of Waterloo, Waterloo, Ontario N2L 3G1, Canada\label{aff22}
\and
Perimeter Institute for Theoretical Physics, Waterloo, Ontario N2L 2Y5, Canada\label{aff23}
\and
INAF-Osservatorio Astronomico di Brera, Via Brera 28, 20122 Milano, Italy\label{aff24}
\and
INFN-Sezione di Genova, Via Dodecaneso 33, 16146, Genova, Italy\label{aff25}
\and
Minnesota Institute for Astrophysics, University of Minnesota, 116 Church St SE, Minneapolis, MN 55455, USA\label{aff26}
\and
Infrared Processing and Analysis Center, California Institute of Technology, Pasadena, CA 91125, USA\label{aff27}
\and
ESAC/ESA, Camino Bajo del Castillo, s/n., Urb. Villafranca del Castillo, 28692 Villanueva de la Ca\~nada, Madrid, Spain\label{aff28}
\and
INAF-Osservatorio di Astrofisica e Scienza dello Spazio di Bologna, Via Piero Gobetti 93/3, 40129 Bologna, Italy\label{aff29}
\and
Dipartimento di Fisica e Astronomia, Universit\`a di Bologna, Via Gobetti 93/2, 40129 Bologna, Italy\label{aff30}
\and
INFN-Sezione di Bologna, Viale Berti Pichat 6/2, 40127 Bologna, Italy\label{aff31}
\and
Institut de Physique Th\'eorique, CEA, CNRS, Universit\'e Paris-Saclay 91191 Gif-sur-Yvette Cedex, France\label{aff32}
\and
Dipartimento di Fisica, Universit\`a di Genova, Via Dodecaneso 33, 16146, Genova, Italy\label{aff33}
\and
Department of Physics "E. Pancini", University Federico II, Via Cinthia 6, 80126, Napoli, Italy\label{aff34}
\and
INAF-Osservatorio Astronomico di Capodimonte, Via Moiariello 16, 80131 Napoli, Italy\label{aff35}
\and
Instituto de Astrof\'isica e Ci\^encias do Espa\c{c}o, Universidade do Porto, CAUP, Rua das Estrelas, PT4150-762 Porto, Portugal\label{aff36}
\and
Faculdade de Ci\^encias da Universidade do Porto, Rua do Campo de Alegre, 4150-007 Porto, Portugal\label{aff37}
\and
European Southern Observatory, Karl-Schwarzschild-Str.~2, 85748 Garching, Germany\label{aff38}
\and
Dipartimento di Fisica, Universit\`a degli Studi di Torino, Via P. Giuria 1, 10125 Torino, Italy\label{aff39}
\and
INFN-Sezione di Torino, Via P. Giuria 1, 10125 Torino, Italy\label{aff40}
\and
INAF-Osservatorio Astrofisico di Torino, Via Osservatorio 20, 10025 Pino Torinese (TO), Italy\label{aff41}
\and
European Space Agency/ESTEC, Keplerlaan 1, 2201 AZ Noordwijk, The Netherlands\label{aff42}
\and
Institute Lorentz, Leiden University, Niels Bohrweg 2, 2333 CA Leiden, The Netherlands\label{aff43}
\and
Leiden Observatory, Leiden University, Einsteinweg 55, 2333 CC Leiden, The Netherlands\label{aff44}
\and
INAF-Osservatorio Astronomico di Roma, Via Frascati 33, 00078 Monteporzio Catone, Italy\label{aff45}
\and
INFN-Sezione di Roma, Piazzale Aldo Moro, 2 - c/o Dipartimento di Fisica, Edificio G. Marconi, 00185 Roma, Italy\label{aff46}
\and
Institute for Theoretical Particle Physics and Cosmology (TTK), RWTH Aachen University, 52056 Aachen, Germany\label{aff47}
\and
INFN section of Naples, Via Cinthia 6, 80126, Napoli, Italy\label{aff48}
\and
Dipartimento di Fisica e Astronomia "Augusto Righi" - Alma Mater Studiorum Universit\`a di Bologna, Viale Berti Pichat 6/2, 40127 Bologna, Italy\label{aff49}
\and
Instituto de Astrof\'{\i}sica de Canarias, V\'{\i}a L\'actea, 38205 La Laguna, Tenerife, Spain\label{aff50}
\and
Institute for Astronomy, University of Edinburgh, Royal Observatory, Blackford Hill, Edinburgh EH9 3HJ, UK\label{aff51}
\and
Jodrell Bank Centre for Astrophysics, Department of Physics and Astronomy, University of Manchester, Oxford Road, Manchester M13 9PL, UK\label{aff52}
\and
European Space Agency/ESRIN, Largo Galileo Galilei 1, 00044 Frascati, Roma, Italy\label{aff53}
\and
Universit\'e Claude Bernard Lyon 1, CNRS/IN2P3, IP2I Lyon, UMR 5822, Villeurbanne, F-69100, France\label{aff54}
\and
Institut de Ci\`{e}ncies del Cosmos (ICCUB), Universitat de Barcelona (IEEC-UB), Mart\'{i} i Franqu\`{e}s 1, 08028 Barcelona, Spain\label{aff55}
\and
Instituci\'o Catalana de Recerca i Estudis Avan\c{c}ats (ICREA), Passeig de Llu\'{\i}s Companys 23, 08010 Barcelona, Spain\label{aff56}
\and
UCB Lyon 1, CNRS/IN2P3, IUF, IP2I Lyon, 4 rue Enrico Fermi, 69622 Villeurbanne, France\label{aff57}
\and
Departamento de F\'isica, Faculdade de Ci\^encias, Universidade de Lisboa, Edif\'icio C8, Campo Grande, PT1749-016 Lisboa, Portugal\label{aff58}
\and
Instituto de Astrof\'isica e Ci\^encias do Espa\c{c}o, Faculdade de Ci\^encias, Universidade de Lisboa, Campo Grande, 1749-016 Lisboa, Portugal\label{aff59}
\and
Department of Astronomy, University of Geneva, ch. d'Ecogia 16, 1290 Versoix, Switzerland\label{aff60}
\and
INAF-Istituto di Astrofisica e Planetologia Spaziali, via del Fosso del Cavaliere, 100, 00100 Roma, Italy\label{aff61}
\and
Aix-Marseille Universit\'e, CNRS/IN2P3, CPPM, Marseille, France\label{aff62}
\and
Universit\'e Paris-Saclay, Universit\'e Paris Cit\'e, CEA, CNRS, AIM, 91191, Gif-sur-Yvette, France\label{aff63}
\and
INFN-Bologna, Via Irnerio 46, 40126 Bologna, Italy\label{aff64}
\and
University Observatory, LMU Faculty of Physics, Scheinerstrasse 1, 81679 Munich, Germany\label{aff65}
\and
Max Planck Institute for Extraterrestrial Physics, Giessenbachstr. 1, 85748 Garching, Germany\label{aff66}
\and
INAF-Osservatorio Astronomico di Padova, Via dell'Osservatorio 5, 35122 Padova, Italy\label{aff67}
\and
Universit\"ats-Sternwarte M\"unchen, Fakult\"at f\"ur Physik, Ludwig-Maximilians-Universit\"at M\"unchen, Scheinerstrasse 1, 81679 M\"unchen, Germany\label{aff68}
\and
Dipartimento di Fisica "Aldo Pontremoli", Universit\`a degli Studi di Milano, Via Celoria 16, 20133 Milano, Italy\label{aff69}
\and
INFN-Sezione di Milano, Via Celoria 16, 20133 Milano, Italy\label{aff70}
\and
Institute of Theoretical Astrophysics, University of Oslo, P.O. Box 1029 Blindern, 0315 Oslo, Norway\label{aff71}
\and
Felix Hormuth Engineering, Goethestr. 17, 69181 Leimen, Germany\label{aff72}
\and
Technical University of Denmark, Elektrovej 327, 2800 Kgs. Lyngby, Denmark\label{aff73}
\and
Cosmic Dawn Center (DAWN), Denmark\label{aff74}
\and
Max-Planck-Institut f\"ur Astronomie, K\"onigstuhl 17, 69117 Heidelberg, Germany\label{aff75}
\and
NASA Goddard Space Flight Center, Greenbelt, MD 20771, USA\label{aff76}
\and
Department of Physics and Astronomy, University College London, Gower Street, London WC1E 6BT, UK\label{aff77}
\and
Department of Physics and Helsinki Institute of Physics, Gustaf H\"allstr\"omin katu 2, University of Helsinki, 00014 Helsinki, Finland\label{aff78}
\and
Universit\'e de Gen\`eve, D\'epartement de Physique Th\'eorique and Centre for Astroparticle Physics, 24 quai Ernest-Ansermet, CH-1211 Gen\`eve 4, Switzerland\label{aff79}
\and
Department of Physics, P.O. Box 64, University of Helsinki, 00014 Helsinki, Finland\label{aff80}
\and
Helsinki Institute of Physics, Gustaf H{\"a}llstr{\"o}min katu 2, University of Helsinki, 00014 Helsinki, Finland\label{aff81}
\and
Laboratoire d'etude de l'Univers et des phenomenes eXtremes, Observatoire de Paris, Universit\'e PSL, Sorbonne Universit\'e, CNRS, 92190 Meudon, France\label{aff82}
\and
SKA Observatory, Jodrell Bank, Lower Withington, Macclesfield, Cheshire SK11 9FT, UK\label{aff83}
\and
Universit\"at Bonn, Argelander-Institut f\"ur Astronomie, Auf dem H\"ugel 71, 53121 Bonn, Germany\label{aff84}
\and
Dipartimento di Fisica e Astronomia "Augusto Righi" - Alma Mater Studiorum Universit\`a di Bologna, via Piero Gobetti 93/2, 40129 Bologna, Italy\label{aff85}
\and
Department of Physics, Institute for Computational Cosmology, Durham University, South Road, Durham, DH1 3LE, UK\label{aff86}
\and
Universit\'e Paris Cit\'e, CNRS, Astroparticule et Cosmologie, 75013 Paris, France\label{aff87}
\and
CNRS-UCB International Research Laboratory, Centre Pierre Bin\'etruy, IRL2007, CPB-IN2P3, Berkeley, USA\label{aff88}
\and
University of Applied Sciences and Arts of Northwestern Switzerland, School of Engineering, 5210 Windisch, Switzerland\label{aff89}
\and
Institut d'Astrophysique de Paris, 98bis Boulevard Arago, 75014, Paris, France\label{aff90}
\and
Institute of Physics, Laboratory of Astrophysics, Ecole Polytechnique F\'ed\'erale de Lausanne (EPFL), Observatoire de Sauverny, 1290 Versoix, Switzerland\label{aff91}
\and
Telespazio UK S.L. for European Space Agency (ESA), Camino bajo del Castillo, s/n, Urbanizacion Villafranca del Castillo, Villanueva de la Ca\~nada, 28692 Madrid, Spain\label{aff92}
\and
Institut de F\'{i}sica d'Altes Energies (IFAE), The Barcelona Institute of Science and Technology, Campus UAB, 08193 Bellaterra (Barcelona), Spain\label{aff93}
\and
DARK, Niels Bohr Institute, University of Copenhagen, Jagtvej 155, 2200 Copenhagen, Denmark\label{aff94}
\and
Space Science Data Center, Italian Space Agency, via del Politecnico snc, 00133 Roma, Italy\label{aff95}
\and
Centre National d'Etudes Spatiales -- Centre spatial de Toulouse, 18 avenue Edouard Belin, 31401 Toulouse Cedex 9, France\label{aff96}
\and
Institute of Space Science, Str. Atomistilor, nr. 409 M\u{a}gurele, Ilfov, 077125, Romania\label{aff97}
\and
Institut f\"ur Theoretische Physik, University of Heidelberg, Philosophenweg 16, 69120 Heidelberg, Germany\label{aff98}
\and
Institut de Recherche en Astrophysique et Plan\'etologie (IRAP), Universit\'e de Toulouse, CNRS, UPS, CNES, 14 Av. Edouard Belin, 31400 Toulouse, France\label{aff99}
\and
Universit\'e St Joseph; Faculty of Sciences, Beirut, Lebanon\label{aff100}
\and
Departamento de F\'isica, FCFM, Universidad de Chile, Blanco Encalada 2008, Santiago, Chile\label{aff101}
\and
Universit\"at Innsbruck, Institut f\"ur Astro- und Teilchenphysik, Technikerstr. 25/8, 6020 Innsbruck, Austria\label{aff102}
\and
Satlantis, University Science Park, Sede Bld 48940, Leioa-Bilbao, Spain\label{aff103}
\and
Instituto de Astrof\'isica e Ci\^encias do Espa\c{c}o, Faculdade de Ci\^encias, Universidade de Lisboa, Tapada da Ajuda, 1349-018 Lisboa, Portugal\label{aff104}
\and
Cosmic Dawn Center (DAWN)\label{aff105}
\and
Niels Bohr Institute, University of Copenhagen, Jagtvej 128, 2200 Copenhagen, Denmark\label{aff106}
\and
Universidad Polit\'ecnica de Cartagena, Departamento de Electr\'onica y Tecnolog\'ia de Computadoras,  Plaza del Hospital 1, 30202 Cartagena, Spain\label{aff107}
\and
Kapteyn Astronomical Institute, University of Groningen, PO Box 800, 9700 AV Groningen, The Netherlands\label{aff108}
\and
Dipartimento di Fisica e Scienze della Terra, Universit\`a degli Studi di Ferrara, Via Giuseppe Saragat 1, 44122 Ferrara, Italy\label{aff109}
\and
Istituto Nazionale di Fisica Nucleare, Sezione di Ferrara, Via Giuseppe Saragat 1, 44122 Ferrara, Italy\label{aff110}
\and
INAF, Istituto di Radioastronomia, Via Piero Gobetti 101, 40129 Bologna, Italy\label{aff111}
\and
Astronomical Observatory of the Autonomous Region of the Aosta Valley (OAVdA), Loc. Lignan 39, I-11020, Nus (Aosta Valley), Italy\label{aff112}
\and
Universit\'e C\^{o}te d'Azur, Observatoire de la C\^{o}te d'Azur, CNRS, Laboratoire Lagrange, Bd de l'Observatoire, CS 34229, 06304 Nice cedex 4, France\label{aff113}
\and
Universit\'e Paris-Saclay, CNRS, Institut d'astrophysique spatiale, 91405, Orsay, France\label{aff114}
\and
School of Physics and Astronomy, Cardiff University, The Parade, Cardiff, CF24 3AA, UK\label{aff115}
\and
Department of Physics, Oxford University, Keble Road, Oxford OX1 3RH, UK\label{aff116}
\and
Aurora Technology for European Space Agency (ESA), Camino bajo del Castillo, s/n, Urbanizacion Villafranca del Castillo, Villanueva de la Ca\~nada, 28692 Madrid, Spain\label{aff117}
\and
ICL, Junia, Universit\'e Catholique de Lille, LITL, 59000 Lille, France\label{aff118}
\and
Instituto de F\'isica Te\'orica UAM-CSIC, Campus de Cantoblanco, 28049 Madrid, Spain\label{aff119}
\and
CERCA/ISO, Department of Physics, Case Western Reserve University, 10900 Euclid Avenue, Cleveland, OH 44106, USA\label{aff120}
\and
Technical University of Munich, TUM School of Natural Sciences, Physics Department, James-Franck-Str.~1, 85748 Garching, Germany\label{aff121}
\and
Max-Planck-Institut f\"ur Astrophysik, Karl-Schwarzschild-Str.~1, 85748 Garching, Germany\label{aff122}
\and
Laboratoire Univers et Th\'eorie, Observatoire de Paris, Universit\'e PSL, Universit\'e Paris Cit\'e, CNRS, 92190 Meudon, France\label{aff123}
\and
Departamento de F{\'\i}sica Fundamental. Universidad de Salamanca. Plaza de la Merced s/n. 37008 Salamanca, Spain\label{aff124}
\and
Universit\'e de Strasbourg, CNRS, Observatoire astronomique de Strasbourg, UMR 7550, 67000 Strasbourg, France\label{aff125}
\and
Center for Data-Driven Discovery, Kavli IPMU (WPI), UTIAS, The University of Tokyo, Kashiwa, Chiba 277-8583, Japan\label{aff126}
\and
Department of Physics \& Astronomy, University of California Irvine, Irvine CA 92697, USA\label{aff127}
\and
Departamento F\'isica Aplicada, Universidad Polit\'ecnica de Cartagena, Campus Muralla del Mar, 30202 Cartagena, Murcia, Spain\label{aff128}
\and
Instituto de F\'isica de Cantabria, Edificio Juan Jord\'a, Avenida de los Castros, 39005 Santander, Spain\label{aff129}
\and
INFN, Sezione di Lecce, Via per Arnesano, CP-193, 73100, Lecce, Italy\label{aff130}
\and
Department of Mathematics and Physics E. De Giorgi, University of Salento, Via per Arnesano, CP-I93, 73100, Lecce, Italy\label{aff131}
\and
INAF-Sezione di Lecce, c/o Dipartimento Matematica e Fisica, Via per Arnesano, 73100, Lecce, Italy\label{aff132}
\and
CEA Saclay, DFR/IRFU, Service d'Astrophysique, Bat. 709, 91191 Gif-sur-Yvette, France\label{aff133}
\and
Institute of Cosmology and Gravitation, University of Portsmouth, Portsmouth PO1 3FX, UK\label{aff134}
\and
Department of Computer Science, Aalto University, PO Box 15400, Espoo, FI-00 076, Finland\label{aff135}
\and
Instituto de Astrof\'\i sica de Canarias, c/ Via Lactea s/n, La Laguna 38200, Spain. Departamento de Astrof\'\i sica de la Universidad de La Laguna, Avda. Francisco Sanchez, La Laguna, 38200, Spain\label{aff136}
\and
Caltech/IPAC, 1200 E. California Blvd., Pasadena, CA 91125, USA\label{aff137}
\and
Universidad de La Laguna, Departamento de Astrof\'{\i}sica, 38206 La Laguna, Tenerife, Spain\label{aff138}
\and
Ruhr University Bochum, Faculty of Physics and Astronomy, Astronomical Institute (AIRUB), German Centre for Cosmological Lensing (GCCL), 44780 Bochum, Germany\label{aff139}
\and
Department of Physics and Astronomy, Vesilinnantie 5, University of Turku, 20014 Turku, Finland\label{aff140}
\and
Serco for European Space Agency (ESA), Camino bajo del Castillo, s/n, Urbanizacion Villafranca del Castillo, Villanueva de la Ca\~nada, 28692 Madrid, Spain\label{aff141}
\and
ARC Centre of Excellence for Dark Matter Particle Physics, Melbourne, Australia\label{aff142}
\and
Centre for Astrophysics \& Supercomputing, Swinburne University of Technology,  Hawthorn, Victoria 3122, Australia\label{aff143}
\and
Department of Physics and Astronomy, University of the Western Cape, Bellville, Cape Town, 7535, South Africa\label{aff144}
\and
DAMTP, Centre for Mathematical Sciences, Wilberforce Road, Cambridge CB3 0WA, UK\label{aff145}
\and
Kavli Institute for Cosmology Cambridge, Madingley Road, Cambridge, CB3 0HA, UK\label{aff146}
\and
Department of Astrophysics, University of Zurich, Winterthurerstrasse 190, 8057 Zurich, Switzerland\label{aff147}
\and
Department of Physics, Centre for Extragalactic Astronomy, Durham University, South Road, Durham, DH1 3LE, UK\label{aff148}
\and
IRFU, CEA, Universit\'e Paris-Saclay 91191 Gif-sur-Yvette Cedex, France\label{aff149}
\and
Univ. Grenoble Alpes, CNRS, Grenoble INP, LPSC-IN2P3, 53, Avenue des Martyrs, 38000, Grenoble, France\label{aff150}
\and
INAF-Osservatorio Astrofisico di Arcetri, Largo E. Fermi 5, 50125, Firenze, Italy\label{aff151}
\and
Dipartimento di Fisica, Sapienza Universit\`a di Roma, Piazzale Aldo Moro 2, 00185 Roma, Italy\label{aff152}
\and
Centro de Astrof\'{\i}sica da Universidade do Porto, Rua das Estrelas, 4150-762 Porto, Portugal\label{aff153}
\and
HE Space for European Space Agency (ESA), Camino bajo del Castillo, s/n, Urbanizacion Villafranca del Castillo, Villanueva de la Ca\~nada, 28692 Madrid, Spain\label{aff154}
\and
INAF - Osservatorio Astronomico d'Abruzzo, Via Maggini, 64100, Teramo, Italy\label{aff155}
\and
Theoretical astrophysics, Department of Physics and Astronomy, Uppsala University, Box 516, 751 37 Uppsala, Sweden\label{aff156}
\and
Institute for Astronomy, University of Hawaii, 2680 Woodlawn Drive, Honolulu, HI 96822, USA\label{aff157}
\and
Mathematical Institute, University of Leiden, Einsteinweg 55, 2333 CA Leiden, The Netherlands\label{aff158}
\and
Institute of Astronomy, University of Cambridge, Madingley Road, Cambridge CB3 0HA, UK\label{aff159}
\and
Univ. Lille, CNRS, Centrale Lille, UMR 9189 CRIStAL, 59000 Lille, France\label{aff160}
\and
Department of Astrophysical Sciences, Peyton Hall, Princeton University, Princeton, NJ 08544, USA\label{aff161}
\and
Space physics and astronomy research unit, University of Oulu, Pentti Kaiteran katu 1, FI-90014 Oulu, Finland\label{aff162}
\and
Center for Computational Astrophysics, Flatiron Institute, 162 5th Avenue, 10010, New York, NY, USA\label{aff163}}

\abstract{We present two extensive sets of 3500$+$1000 simulations of dark matter haloes
  on the past light cone, and two corresponding sets of simulated (`mock') galaxy
  catalogues that represent the \Euclid spectroscopic sample. The simulations were
  produced with the latest version of the {\pin} code, and provide the largest, public set
  of simulated skies. Mock galaxy catalogues were obtained by populating haloes with
  galaxies using an halo occupation distribution (HOD) model extracted from the Flagship
  galaxy catalogue provided by Euclid Collaboration. The Geppetto set of 3500 simulated
  skies was obtained by tiling a $1.2\,\hgpc$ box to cover a light-cone whose sky
  footprint is a circle of {\tdeg} radius, for an area of 2763 \sqdeg and a minimum halo
  mass of $1.5\times10^{11}$\,\hmsun. The relatively small box size makes this set unfit
  for measuring very large scales. The EuclidLargeBox set consists of 1000 simulations of
  $3.38\,\hgpc$, with the same mass resolution and a footprint that covers half of the
  sky, excluding the Milky Way zone of avoidance. From this we produced a set of 1000
  EuclidLargeMocks on the {\tdeg} radius footprint, whose comoving volume is fully
  contained in the simulation box. We validated the two sets of catalogues by analysing
  number densities, power spectra, and 2-point correlation functions, showing that the
  Flagship spectroscopic catalogue is consistent with being one of the realisations of the
  simulated sets, although we noticed small deviations limited to the quadrupole at
  $k>0.2\,\kmpc$. We show cosmological parameter inference from these catalogues and
  demonstrate that using one realisation of EuclidLargeMocks in place of the Flagship mock
  produces the same posteriors, to within the expected shift given by sample variance.
  These simulated skies will be used for the galaxy clustering analysis of \Euclid's Data
  Release 1 (DR1), and an even larger set of simulations is planned for the next releases.}

\keywords{Methods: numerical; Surveys; Cosmology: observations; Large-scale structure of
  Universe; Cosmological parameters}

   \titlerunning{Thousands of \Euclid skies}
   \authorrunning{Euclid Collaboration: P.~Monaco et al.}
   
   \maketitle
   
\section{Introduction}
\label{sec:intro}

The \Euclid satellite \citep{EuclidSkyOverview}, launched on July 1st, 2023, is mapping
the visible Universe to redshifts of at least $z=2$ with the aim of shedding light on its
dark sector. Together with other Stage IV surveys, like DESI \citep{DESI}, Rubin--LSST
\citep{Rubin}, Roman \citep{Roman}, SphereX \citep{SphereX}, or the SKAO cosmological
surveys \citep{SKAO}, \Euclid is starting to contribute a cosmological data flood, opening
to the possibility of constraining cosmology at an unprecedented precision level,
comparable to what the CMB provides at $z\sim 1100$.

In particular, \Euclid is surveying the sky with the VIS visible imager
\citep{EuclidSkyVIS}, designed for weak lensing studies, and the near-infrared imager and
spectrometer \citep[NISP,][]{EuclidSkyNISP} that uses slitless spectroscopy to measure
redshifts for a sample of emission-line galaxies; these redshifts are the basis for galaxy
clustering measurements, that is the focus of our research. The wavelength range of the
`red grism' used by NISP to disperse the light makes it possible to cover the {\ha} line
in emission in a redshift range of roughly $z\in[0.9,1.8]$. We expect that most galaxies
with reliable redshift have line flux \fha larger than a fiducial value of
$f_0=2\times10^{-16}$ \flux. The Euclid wide survey \citep[EWS,][]{Scaramella-EP1} will
cover $\sim$\,14\,000 \sqdeg of the sky, excluding the zone of avoidance of the Milky Way
and the ecliptic plane where zodiacal light hampers deep observations. The EWS will be
sided by the Euclid deep survey (EDS), consisting of 53 \sqdeg in three disconnected
regions; these fields will be spectroscopically observed 15 times with the red grism, plus
25 times using a `blue grism' to extend the spectral coverage to shorter wavelengths. The
EDS will provide a highly pure and complete version of the galaxy sample detected in the
EWS, allowing a precise assessment of its purity and completeness, a crucial ingredient to
control systematic effects.

The sheer amount of high-quality data is already challenging the traditional galaxy survey
processing methods. Access to a large cosmic volume, sampled with a dense galaxy
catalogue, will allow us to beat down the statistical uncertainty of the measurement, so
the error budget will be dominated by systematic effects. Because galaxy clustering is
based on counting galaxies down to the faintest accessible flux, control of the depth of
the survey will be crucial to have an unbiased estimate of the galaxy density. Because
most redshift measurements will be based on spectra with low signal-to-noise ratio,
control of redshift errors (and the consequent emergence of interlopers where redshifts
are catastrophically wrong) will be key to keeping the purity of the spectroscopic sample
high. Eventually, in a standard likelihood approach, we will compress data into a summary
statistics of galaxy clustering, like the 2-point correlation function in configuration or
Fourier space (the power spectrum), possibly augmented with the corresponding 3-point
functions, and we will compare these with a model that predicts the summary statistics as
a function of cosmological and nuisance parameters. The difference between measurement and
model must be related to a covariance matrix, which for its cosmological part is a
higher-order moment of clustering. The covariance matrix should also contain contributions
from systematics error, representing the uncertainty in the mitigation of known systematic
effects. As shown by \cite{Colavincenzo2017}, even with the simplest assumptions, where
the correction to the galaxy density is multiplicative, the cosmological covariance cannot
be simply separated as a sum of contributions from cosmology and systematic effects. A
numerical characterisation of the covariance matrix, obtained by processing thousands of
simulated skies, remains the most effective way to address this challenge.

In this paper we present the simulated spectroscopic skies that have been produced by the
Euclid Collaboration to face these challenges. These are based on the largest set of
simulated dark matter haloes on the past light cone ever produced. We discuss two sets of
simulated \Euclid spectroscopic skies (whose properties are reported in
Table~\ref{table:simulations}), performed with the {\pin} code \citep{Monaco2002} for a
flat $\Lambda$ cold dark matter ($\Lambda$CDM) model compatible with {\it Planck} results
\citep{Planck2018}. The first set of 3500 realisations, named Geppetto, is based on a
relatively small box of $1.2\,\hgpc$, sampled with $2160^3$ particles to achieve a minimum
halo mass of $1.5\times10^{11}\,\hmsun$. Each light-cone covers a circle of 30 degrees of
radius on the sky, covering 2763 {\sqdeg}, a bit larger than the area planned to be
covered by \Euclid's DR1 \citep{Scaramella-EP1},\footnote{
In the original plan, DR1 consisted of two connected areas on the North and South 
ecliptic sky, with a rather complicated shape. The \tdeg circle 
footprint would cover 99\% of the North survey and 95\% of the South survey. However, DR1 is 
being re-designed, so we do not present a detailed assessment here. 
} and extends to redshift $z=2$. The second
set of 1000 realisations (named EuclidLargeBox) is based on a $3.38\,\hgpc$ box, sampled
by $6144^3$ particles and achieving the same mass resolution as Geppetto. The light-cones
cover half of the sky, excluding a zone of avoidance of the Milky Way, for a semi-aperture
angle of $70$ deg, and extend to $z=4$.

The paper is organised as follows. Section~\ref{sec:context} expands on the context and on
the motivation for the production of these simulations. Section~\ref{sec:code} describes
the improvements to the {\pin} code that were required to scale up to this large number of
particles. Section~\ref{sec:simulations} describes in detail the simulations produced,
while Sect.~\ref{sec:hod} describes the techniques adopted to populate haloes with
galaxies and produce mock galaxy catalogues; details in the adopted matching of {\pin} and
$N$-body masses are reported in an Appendix. Section~\ref{sec:validation} presents a set
of measurements performed to validate the sets of mock catalogues and demonstrate their
utility to obtain numerical covariance matrices for \Euclid. Finally,
Sect.~\ref{sec:conclusions} gives the conclusions.

\section{Context}
\label{sec:context}

Our starting point is the upcoming spectroscopic sample of \Euclid
\citep{EuclidSkyOverview}, consisting of $\sim$\,30 million galaxies with reliable
redshift measurement in the range $0.9<z<1.8$ over 14\,000 \sqdeg. Systematic effects for
the measurement of the galaxy density are represented by constructing a `visibility mask',
a function of sky position and redshift that quantifies, to the best of our knowledge, the
probability that a galaxy gets a reliable redshift measurement given its properties and
the local observing conditions. This visibility mask is applied to a dense set of
unclustered galaxies to obtain a `random catalogue', that is used as a reference to
compute the galaxy density contrast, thus mitigating systematic effects induced by a
non-uniform selection probability. All estimators of summary statistics that are based on
the galaxy density rely on this random catalogue.

To compute the numerical covariance of a summary statistics we start from a set of mock
catalogues that represent galaxies in the past light cone, on an angular footprint
matching the \Euclid survey. Selection biases can be imposed to mock catalogues by
applying the same visibility mask used to construct the random catalogue, including
catastrophic redshift errors. Then the same pipeline used to process the real data
catalogue is applied to the mock catalogues. The resulting measurements are combined to
produce a numerical covariance, which is then fed to a likelihood code to estimate
cosmological parameters. This approach has been used in several surveys
\citep[e.g.][]{Manera2013,Balaguera2023}, and has the clear advantage of being realistic
in representing the level of non-linearity of the density field, the complexity of galaxy
bias and the treatment of systematic effects .

A numerical covariance matrix is affected by noise as it is sampled by a finite number of
mocks; for \Euclid, the requirement on its accuracy is that parameter uncertainties (for
the final sample after the nominal six years of the mission) should vary by less than
10\%, with respect to an infinite number of mocks. In Euclid Collaboration: Sanchez et al.
(in prep.), we estimate that 3500 mocks are needed to have an accurate, brute-force
numerical covariance matrix. This number is mostly determined by the total number of
measurements for which the covariance is computed and by the accuracy requirement. That
paper also presents tests of several methods to {\it de-noise} a numerical covariance
matrix, demonstrating that $\sim$\,100 mocks may be sufficient to achieve good accuracy
\citep[see also][]{Fumagalli2022} when addressing the relatively limited data vector of
2-point correlation functions (in configuration or Fourier space). However, this figure
may be optimistic: the inclusion of 3-point statistics will further increase the length of
the data vector, moreover a precise characterisation of systematic effects requires to
analyse a large number of mock catalogues (e.g. \citealt{EP-Risso}; 
Euclid Collaboration: Lee et al., in prep.). Under these premises, we consider 1000
as a minimal requirement on the number of realisations for a set of simulated catalogues
for \Euclid.

A single simulation should ideally cover a volume that can include the largest connected
patch of the surveyed volume, so as to properly represent super-sample covariance; this
pushes requirements on the box size beyond $3\,\hgpc$. The simulation should also resolve
the smallest halo that contains an observed galaxy, that (as argued in
Sect.~\ref{sec:hod}) is expected to be $\sim$\,$2\times 10^{11}$\,\hmsun for the
spectroscopic sample; this pushes requirements on particle mass towards $10^9$\,\hmsun.
One such simulation was used to create the Flagship galaxy mock catalogue
\citep{EuclidSkyFlagship}, a comprehensive catalogue representing galaxies in all aspects
that are of interest for \Euclid, including weak lensing, galaxy clusters, and galaxy
evolution. The single Flagship numerical simulation used to create the galaxy mock
catalogue costed $\sim$\,1\,000\,000 node-hours (corresponding to 68\,000\,000 core hours)
on 4000 nodes of Piz Daint supercomputer, and required storage of order of petabytes.
Running thousands of such simulations is out of the question.

An alternative to the $N$-body approach is to run fast simulations using approximate
methods, to obtain thousands of catalogues of dark matter haloes at the cost of one single
equivalent simulation. Here speed is obtained by treating the evolution of perturbations
in the mildly non-linear regime using some flavour of Lagrangian Perturbation Theory
(LPT), avoiding to integrate the trajectories of particles within dark matter haloes.
Approximate methods were reviewed by \cite{Monaco2016}. They can be broadly separated into
two classes, `predictive' and `calibrated'. The first class is typically based on
Lagrangian methods that range from semi-analytic, peak-based methods like
\texttt{Peak-Patch} \citep{Bond1996,Stein2019} or {\pin} to fast $N$-body codes, typically
particle-mesh (PM) codes, like \texttt{FastPM} \citep{Feng2016} or \texttt{COLA}
\citep{Tassev2013}. The predictive nature of these codes implies that once they are
properly calibrated on $N$-body simulations they can be applied to any box size, mass
resolution, and cosmology; as a matter of fact their computational cost, at fixed box size
and mass resolution, is broadly proportional to the accuracy they can achieve. The second
class is typically based on a sophisticated bias scheme applied to a low-resolution
density field, like \texttt{EZMOCKS} \citep{Chuang2015}, \texttt{Patchy}
\citep{Kitaura2015}, \texttt{BAM} \citep{Balaguera2019}, \texttt{COVMOS}
\citep{Baratta2023}, \texttt{PineTree} \citep{Ding2024}, or \texttt{GOTHAM}
\citep{Pandey2024}. These methods must be calibrated on an $N$-body simulation each time
they have to be used, but the time needed to produce a single realisation is tiny, even
when compared with \pin, and so they are useful for a massive production of tens of
thousands of mocks.

Approximate methods were thoroughly tested in \cite{Nifty2015} and in a series of studies
in preparation of {\Euclid} \citep{Lippich2019,Blot2019,Colavincenzo2019}. In these three
papers 300 $N$-body simulations were compared with several approximate methods run on the
same initial conditions, addressing respectively the 2-point correlation function, the
power spectrum, and the bispectrum of dark matter haloes at various mass thresholds.
Consistency was tested by using these covariances for parameter estimation and requiring
that the relative increase of parameter uncertainties is less than 10\%. As a result,
while the average of the summary statistics may be biased, cosmological parameters
inferred using the numerical covariance obtained from $N$-body were found consistent, for
many methods, with those obtained using approximate simulations, with differences in
parameter uncertainties within the requirement mentioned above.

To produce the simulations presented in this paper, we used the 5.0 version (V5) of {\pin}
\citep{Monaco2002,Munari2017}.\footnote{\url{https://github.com/pigimonaco/Pinocchio}} The
code is described in detail in Sect.~\ref{sec:code}, we just outline here the motivation
for its use. As pointed out in \cite{Munari2017b}, a predictive approximate method must
solve two problems: first compute how particles move from their Lagrangian position,
second determine how particles group into haloes. While the first problem is relatively
easy to solve using LPT or a PM code, the second problem in principle requires accurate
particle positions at scales smaller than the virial radius of the halo, making it
challenging even for fast PM codes to reconstruct small haloes. {\pin} can be seen as a
(semi-analytic) halo finder in Lagrangian space, thus solving the difficult problem of
associating particles to haloes even with rather approximate displacements. This makes it
easier to LPT to represent the density field; for instance, \citet{Munari2017b} showed
that, when multi-stream regions are correctly collapsed into haloes, 3LPT yields an
increase in accuracy in halo positions with respect to 2LPT that is as large as the
increase in accuracy going from the Zeldovich approximation to 2LPT. This is not true when
straight LPT is applied to initial conditions, because higher LPT orders lead to stronger
dispersion of particles after orbit crossing, thus losing the advantage of higher order.
{\pin} has been extensively used in the literature; for instance, \cite{Oddo2020,Oddo2021}
and \cite{Rizzo2023} used 10\,000 realisations of the same setting of \cite{Lippich2019}
to check that {\pin} is correctly representing the covariance of the bispectrum of dark
matter haloes.

\section{Code}
\label{sec:code}

\pin, a C code with \texttt{MPI} parallelization, has been designed to generate very good
approximations of catalogues of dark matter haloes, both in periodic boxes and in the past
light cone, with full information on their mass, position and merger history. Using
excursion set theory and ellipsoidal collapse, the code computes for each particle an
estimate of the time of orbit crossing (the collapse time), then groups the collapsed
particles into haloes, following their merger history, and places haloes at the final
position using 3LPT. The original code is described in \cite{Monaco2002}, the 2LPT and
3LPT extensions in \cite{Monaco2013} and \cite{Munari2017}. This section describes the
latest technical developments of the code.

The starting point is the realisation of a linear density field on a regular grid, as in
the generation of the initial conditions of a cosmological $N$-body simulation. The code
is made of two main parts, the computation of collapse times and LPT displacements for
each particle, and the grouping of collapsed particles into haloes (`fragmentation'), with
the construction of halo merger histories and a light-cone with continuous time sampling.
Collapse times are computed by Gaussian-smoothing the linear density field on many
smoothing radii, then computing the second derivatives of the potential with fast Fourier
transforms (FFTs); these are used to compute the collapse redshift of each particle using
ellipsoidal collapse. We define the inverse collapse time as $F=1+z_{\rm c}$, and store
its highest value $F_{\rm max}$ for all smoothing radii. At the final smoothing radius
$R=0$ (meaning that the variance of the linear density field is only limited by the
Lagrangian grid), the LPT displacement fields are computed, amounting to four vectors for
each particle (of the three 3LPT displacement fields we only compute the first two, the
third rotational term being negligible). The second part of the code uses the collapse
times and displacements to group particles into haloes, with an algorithm that mimics
hierarchical clustering. Because collapse here is identified with orbit crossing,
collapsed particles are not necessarily contained in haloes, they may be part of the
filamentary network that joins haloes. So particles may be classified into uncollapsed
(still in single-stream regime), filaments, and halo particles. Having recognised the
haloes without really running the simulation (we just perform a single 3LPT time-step when
needed), we can see \pin as a halo finder that works on the Lagrangian space of initial
conditions, plus a 3LPT engine to place the haloes at the right position.

The fourth version (V4) of \pin is described in \cite{Munari2017}, where scaling
properties are demonstrated to be nearly ideal for a range of conditions, culminating in
boxes sampled with $2160^3$ particles as the first Millennium simulation
\citep{Millennium}. Parallelisation is achieved with the MPI (message passing interface)
protocol. However, scaling to larger sizes is hampered by two factors. Firstly, the FFT
library \texttt{FFTW} \citep{FFTW} distributes the 3D computational domain in planes, so
each task must allocate memory for as many particles as a single plane. For a memory need
of 350 bytes per particle, a plane of a $6144^3$ box would require 12 GB, making it hard
to run on all the available cores in a node. Secondly, fragmentation is an intrinsically
scalar process, so parallelisation is achieved by dividing the computation domain in
sub-boxes, chosen to present the smallest surface over volume; this implies a round of
communications among tasks to move from the FFT domain to the sub-box domain, that has a
small computational cost. However, haloes at the border of the domain will not be
correctly reproduced, unless fragmentation is performed on an augmented domain that
contains all the particles that are needed to properly produce haloes. These `ghost
regions', called boundary layer in the code jargon, create an overhead in memory and in
computing time. They cannot be arbitrarily small as their size must be as large as the
Lagrangian size of the largest halo that is predicted to be present in the box. At $z=0$
this can amount to $\sim$\,35\,\hmpc, placing a hard limit on the number of tasks the code
can be distributed over.

\subsection{FFT solver}

The first limitation was overtaken by using the \texttt{pFFT}
library.\footnote{\url{https://github.com/mpip/pfft}} This uses \texttt{FFTW} to perform
1D Fourier transforms, but wraps the 2D and 3D transforms differently, distributing memory
in planes, pencils or sub-volumes. When memory constraints allow it, distribution in
planes is the most convenient one, while distribution in pencils and in sub-volumes has a
slightly increasing computational cost. While usage of \texttt{pFFT} simply required an
adaptation of the code that performs FFTs, the code that creates the initial conditions
required a deep re-design. The fourth version of \pin implemented the same loop in
$k$-space present in \texttt{N-GenIC} and \texttt{2lpt-IC} codes \citep{Springel2005,
  Crocce2006}, making it possible to reproduce simulations started from initial conditions
produced with those codes by just providing the same seed for random number generator.
\texttt{N-GenIC} works as follows: each task creates a 2D table of random seeds, one for
each row of cells along the box, then loops in $k$-space by drawing seeds consecutively
from the random sequence relative to each pencil, assuring that the same result is
achieved for any domain decomposition. Moreover, $k$-space is populated starting from the
fundamental mode and progressing in concentric cubes in Fourier space. This way, doubling
the number of grid points per side will result in a higher resolution realisation that has
exactly the same larger-scale modes as the lower resolution one. This feature is
maintained in the current version of the code; detailed technical aspects will be
presented in a forthcoming technical publication.

\subsection{Fragmentation code}

The second limitation to code scaling concerns the construction of haloes from collapsed
particle (fragmentation), and it was overtaken by changing the way the code stores and
accesses memory. The first part of the code, that computes collapse times, produces for
each particle its inverse collapse time $F_{\rm max}$ and the four LPT displacement
fields. In the sub-box domain, the tasks do not really need these products for all
particles, because particles that do not collapse by the final redshift will never be
processed; moreover, not all the regions of the boundary layer are necessarily of
interest. The code was then redesigned to have a more accurate selection of the particles
that are loaded in the sub-box domain. To achieve this aim, each task gathers the
properties of all collapsed particles in its `well-resolved' region, that is the patch of
Lagrangian space assigned to the task, augmented just by a 1-particle ghost region to
properly compute the peaks of $F_{\rm max}$. This is done by creating a Boolean map of the
whole computational domain (each point represented by a bit), setting to true the bits
corresponding to the particles of interest. Of this domain, the task will receive (through
a hypercubic communication scheme among tasks) only the products of the particles that are
predicted to collapse by the end of the simulation. After that, the fragmentation code is
run in a minimal configuration (no light-cone construction, no output) with the available
information. At the end of this first fragmentation, haloes near the borders are not
reconstructed properly but we know where they are. Another all-to-all communication round
is then called: each task creates a second Boolean map that is set to \texttt{true} for
all the particles that lie within a certain distance from the haloes and are beyond the
well-resolved region. This distance is set as the Lagrangian radius of the halo (in grid
units this is the cubic root of the number of its particles) times a boundary layer factor
$f_{\rm bl}\sim2$--3. After the products of these particles are communicated, the
fragmentation is started again.

This implementation limits memory overhead by a large factor, improving the scaling of the
fragmentation part of the code. Another large factor in memory requirement is obtained by
storing the products of the collapse time part in single floating-point precision, while
most of the code works in double precision; it has been checked that this creates an
acceptable decrease in numerical accuracy, while yielding a drastic advantage in memory
requirements. As a result, we pass from 350 to 150 bytes per particle as a minimum needed
by 3LPT. However, while the previous version of the code had a predictable mapping between
sub-box space domain and memory, accessing the memory location of a particle is less
straightforward, and it is made possible by a series of pointers and some bisector
searches. This increases the computing time in ideal cases, but the improvement in scaling
overcomes this problem.

In this implementation, each task requires a variable amount of memory; for instance, if
the largest cluster in the box falls at the border of a domain, its task will require a
larger number of particles from the boundary layer, while a task with voids at its border
will have a smaller overhead. This imbalance is handled by allowing the same overhead to
all tasks; if a task fills all of its memory it issues a warning but does not halt the
computation, and the largest overhead needed by the task is output at the end of the code,
so it is easy to achieve an optimisation of memory usage after few test runs. This is
considered as an acceptable strategy to prepare the production of thousands of massive
runs.

\section{Dark matter catalogues and light-cones}
\label{sec:simulations}

As discussed above, we produced two sets of simulations aimed at achieving a large number
of realizations based on a small box (Geppetto) and a large box that can contain a survey,
with a more limited number of realizations (EuclidLargeBox). The main properties of these
sets are reported in Table~\ref{table:simulations}. Both sets are based on a $\Lambda$CDM
cosmology with parameters close to \cite{Planck2018}: $\Omega_{\rm m}=0.32$,
$\Omega_\Lambda=0.68$, $\Omega_{\rm b}=0.049$, $h=0.67$, $\sigma_8=0.83$, $n_{\rm
  s}=0.96$. These parameters are consistent with Flagship 2 simulation and mock galaxy
catalogue; in the Flagship 2 simulation we have $\Omega_{\rm m}=0.319$ and massive
neutrinos, but these differences have no practical effect for the present purposes. For
each simulation we produced a few dark matter halo catalogues in periodic boxes and a
complete catalogue in the past light cone, with a sky footprint consisting of a circle
with a given radius. Moreover, merger trees are available for all haloes. The minimum halo
mass is set to 10 particles; this would be an unacceptably low value for an $N$-body
simulation, but it is acceptable for a semi-analytic code where haloes are related to
peaks of the inverse collapse time and each halo starts with one particle.

\begin{table*}
\caption{Main properties of the mock catalogues.}
\begin{center}
\begin{tabular}{l|ccccccccccc}
Name & $L_{\rm s}$ & $N_{\rm part}$ & $N_{\rm real}$ & min. $M_{\rm h}$ & $V_{\rm s}$ & tot. $V_{\rm s}$ & $\theta$ & area & $z_{\rm start}$ & $V_{\rm lc}$  & tot. $V_{\rm lc}$ \\ 
     & {\small\hgpc}   &              &              &  {\small\hmsun}  & {\small\cgpc} & {\small\cgpc}   &          & {\small\sqdeg} &           & {\small\cgpc}      & {\small\cgpc}  \\
\hline
Geppetto & 1.2 & $2160^3$ & 3500 & $1.52\times10^{11}$ &  1.73 & 6047 & $30^\circ$ & 2763 &  2.0 & 12.72  & 44\,529 \\
EuclidLargeBox & 3.38 & $6144^3$ & 1000 & $1.48\times10^{11}$ & 38.61 & 38\,614 & $70^\circ$ & 13\,572 &  4.0 & 163.86  & 163\,862 \\
\hline
Minerva-like & 1.5 & $1000^3$ & 10\,000 & $2.67\times10^{12}$ &  3.38 & 33\,750 & --- & --- & --- & --- & --- \\
NewClusterMocks & 3.87 & $2160^3$ & 1000 & $4.90\times10^{12}$ & 57.96 & 57\,960 & $60^\circ$ & 10\,313 &  2.5 & 69.70  & 69\,700 \\
\hline
\end{tabular}
\label{table:simulations}
\tablefoot{We report here
  Geppetto and EuclidLargeBox, together with the Minerva-like set used in \cite{Oddo2020}
  and NewClusterMocks used in \cite{Fumagalli21}. Columns give the simulation box side
  $L_{\rm s}$, the number of particles in the simulation $N_{\rm part}$, the number of
  realisations $N_{\rm real}$, the minimum halo mass $M_{\rm h}$, the simulation volume
  $V_{\rm s}$ and its total over the realisations, the semi-aperture of the light-cone
  $\theta$, the survey area, the starting redshift of the light-cone $z_{\rm start}$, the
  volume of a single light-cone $V_{\rm lc}$ and its total over the realisations.}
\end{center}
\end{table*}

\subsection{Geppetto simulations}
\label{sec:Geppetto}

The first set of $N_{\rm real}=3500$ simulations, called Geppetto, was designed to be
massive but relatively inexpensive, to produce a brute-force numerical covariance for the
target sample of {\ha} emitters. It is based on a relatively small box of side $L_{\rm
  s}=1200$\,\hmpc, sampled with $N_{\rm part}=2160^3$ particles. With the given cosmology,
the particle mass is $1.52\times10^{10}$\,\hmsun, and the smallest halo has a mass of 10
particles or $M_{\rm h}=1.52\times10^{11}$\,\hmsun. The light-cones start at $z_{\rm
  start}=2$ and cover a circle of radius of $\theta=30^\circ$, for a total area of
$2\pi\,(1-\cos\theta)$ steradians, equal to 2763 \sqdeg.\footnote{
Noticeably, this specific sky footprint implies that the comoving volume of the light-cone
is a 3D cone itself; however, the word `cone' in the two cases refers to a spacetime and
to a volume in 3D space, thus assuming a pretty different meaning.
} Outputs at fixed time are available at redshifts 2, 1.8, 1.5, 1.35, 1, 0.9, 0.5, and 0.

These simulations were mostly run on the Pleiadi system of INAF (`Istituto Nazionale di
Astrofisica'), a Tier-2 facility with 66 computing nodes, each with 36 cores and 256 GB or
RAM. Each run required 20 nodes, and lasted around 17 minutes, for a cost of 200 core-h
per run. The total cost amounted to only 700\,000 core-h. Each run required 23 GB of
storage, for a total need of almost 80 TB. As a matter of fact, the first 600 realisations
were run with V4.1.3 of the code, with minimal changes (using single precision of the
products of collapse time calculation and using a rather limited boundary layer to
minimise memory overhead) to allow the code to scale to the needed mass resolution.

\begin{figure}
\centering
\includegraphics[width=\hsize]{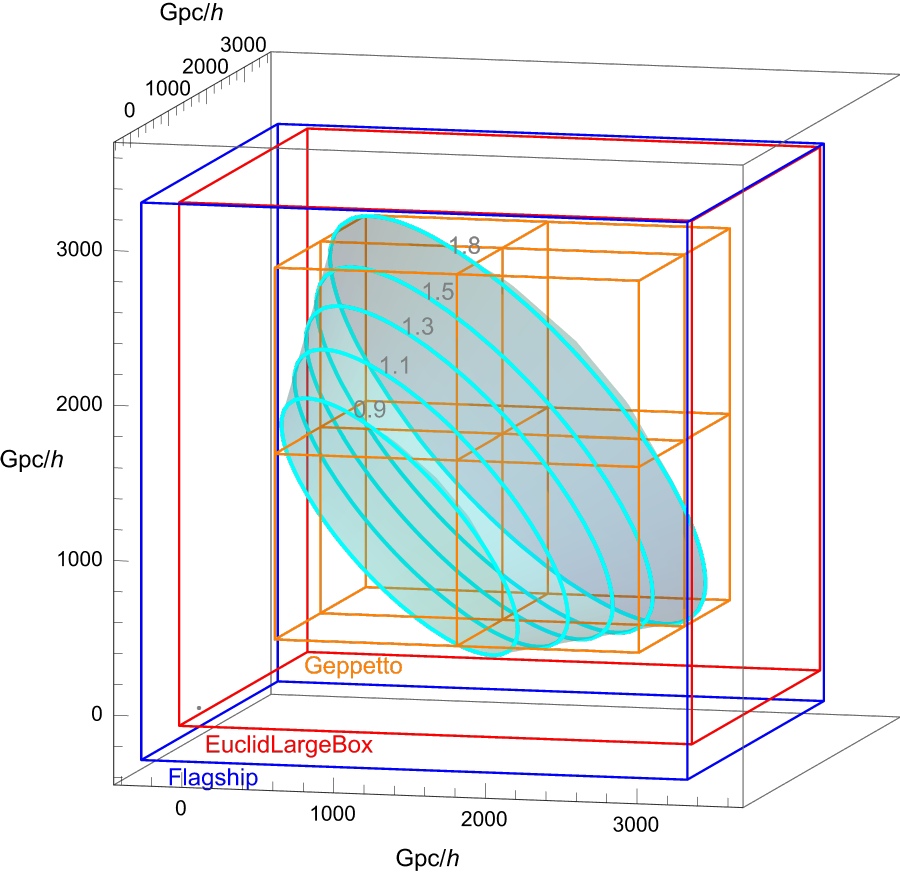}
\caption{Graphical representation of the comoving volume of the {\tdeg} radius survey
  footprint, highlighting the four redshift bins bounded by redshifts
  $[0.9,1.1,1.3,1.5,1.8]$, immersed in the comoving volume of the Flagship simulation
  (that is our reference), of EuclidLargeBox, and of Geppetto simulation boxes. The
  Geppetto box is tiled a few times to cover the survey volume; for sake of clarity, the
  tiling is limited to the main replications. Here the central axis of the footprint lies
  on the main diagonal of the box, however all EuclidLargeBox and most Geppetto boxes have
  random orientations of this axis.}
\label{fig:replicas}
\end{figure}

\begin{figure}
\centering
\includegraphics[width=\hsize]{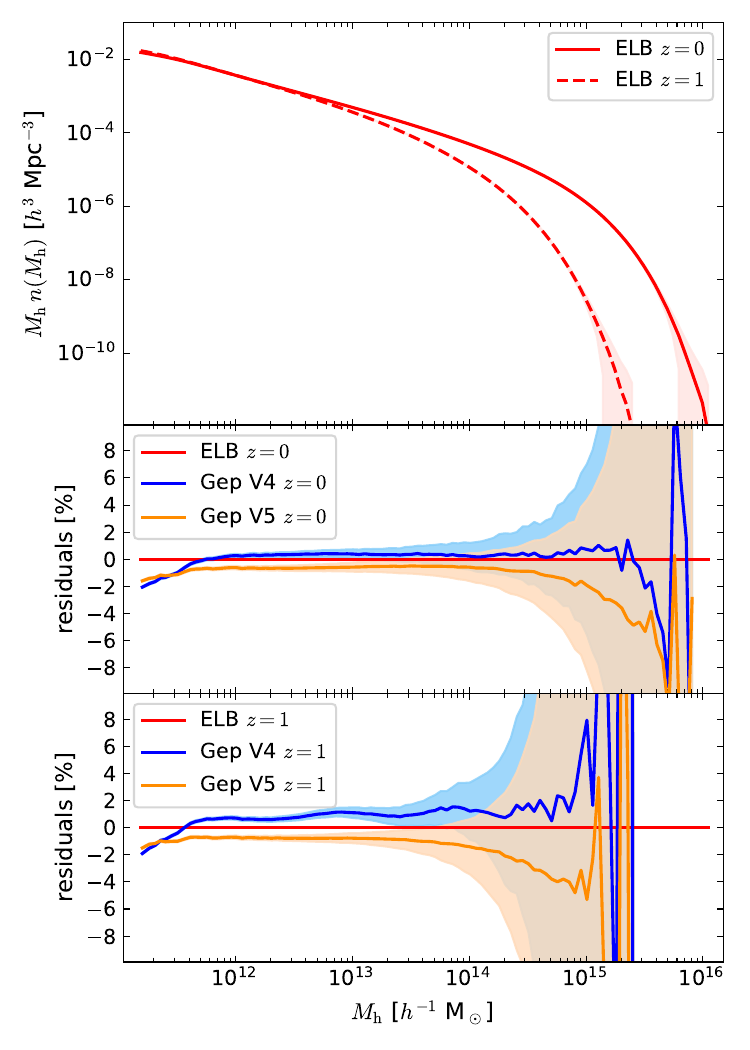}
\caption{Halo mass function $n(M_{\rm h})$ of the simulated sets. The upper panel only
  shows the mass function of EuclidLargeBox (ELB) for $z=0$ and $z=1$, where the line
  gives the average over the 1000 simulations and the shaded area gives its sample
  variance. The two lower panels show relative residuals, $n/n_{\rm elb}-1$ in percent, of
  the mass function of the first 600 Geppetto simulations, run with V4 of the code (Gep
  V4), and the other 2900, run with V5 (Gep V5), with respect to the EuclidLargeBox
  measurement, respectively at $z=0$ (mid panel) and $z=1$ (lower panel). Here the
  reported sample variance is relative only to the Geppetto sets.}
\label{fig:haloMF}
\end{figure}

The comoving volume of the produced light-cones is 12.72\,\cgpc, that is 7.35 times larger
than the simulated volume, 1.73\,\cgpc. Figure~\ref{fig:replicas} reports the light-cone
volume, limited to the four redshift bins that contain the spectroscopic sample (defined
below). This is immersed in the Flagship simulation box (46.7\,\cgpc) that completely
contains it. The figure also reports the Geppetto box, tiled to cover the volume. For sake
of clarity we only report the main replications that cover most of the survey volume, in
fact the tiling extends to all replications that have a non-null intersection with the
cone.\footnote
{The code includes an algorithm dedicated to determining whether a given cone and cube
  intersect, a geometrically non-trivial problem despite involving only high-school-level
  maths.}
Tiling is applied by using periodic boundary conditions to guarantee continuity of the
density field. To minimise the statistical effect of these replications, the orientation
of the light-cone volume is varied: while in the first 700 realisations the central axis
of the cone, pointing from the observer to the centre of the sky footprint, is directed
along the main diagonal of the box as in Fig.~\ref{fig:replicas}, in the other cases the
axis is aligned along a random direction.

Averaging over so many realisations produces a very smooth halo mass function that beats
down sample variance to negligible levels up to the mass of galaxy groups. The halo mass
function is shown in Fig.~\ref{fig:haloMF}; we separate the Geppetto mocks in two sets,
the first 600 simulations, performed with V4 of the code, and the remaining 2900, run with
V5. We also include results for the EuclidLargeBox set described below. The upper panel
gives the halo mass function $M_{\rm h}\,n(M_{\rm h})$ at $z=0$ and $z=1$, and because the
three curves are indistinguishable we only report results for the EuclidLargeBox set. The
lower panels show the relative difference of the Geppetto halo mass functions with respect
to the EuclidLargeBox one, $n/n_{\rm elb}-1$ in percent, again at $z=0$ (mid panel) and
$z=1$ (lower panel). In all cases the lines show the average mass functions while the
shaded areas give their sample variance; in the residuals the variance refers only to the
Geppetto sets, for the denominator $n_{\rm elb}$ we only use the average. The comparison
of the results of V4 and V5 versions shows subtle percent-level differences. This
discrepancy represents a residual of the calibration process, which is accounted for and
reabsorbed during the subsequent HOD calibration (as discussed in Sect.~\ref{sec:hod}).

\begin{figure} 
\centering
\includegraphics[width=\hsize]{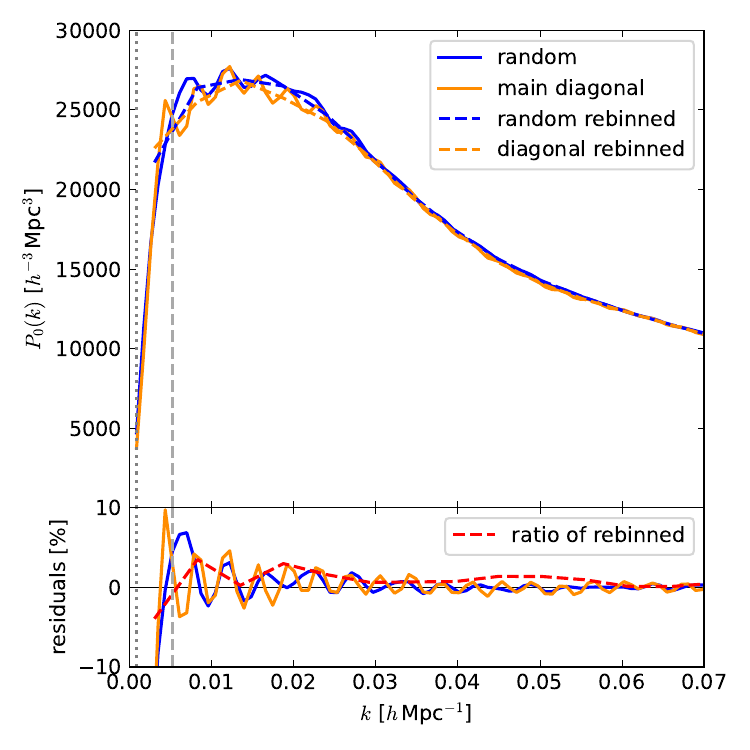}
\caption{Power spectrum monopole of {\pin} dark matter haloes from the Geppetto
  light-cones, with $M_{\rm h}>10^{12}$\,\hmsun in the redshift range $z\in[0.9,1.1]$. The
  continuous lines show measurements in a box of $L_{\rm m}=7.2\,\hgpc$ sampled on its
  fundamental frequency $k_{\rm fm}$, the dashed lines show the same measurements with a
  sampling of $6k_{\rm fm}$, equal to the fundamental mode of the simulation box $k_{\rm
    fs}$. Blue and orange lines show respectively averages over 100 mocks with random
  orientations of the cone axis and with axis aligned with the main diagonal of the
  simulation box. The lower panel shows the residuals of the measurements along a random
  orientation with respect to those with axis along the main diagonal, plus, in red, the
  residuals of the two rebinned measurements.}
\label{fig:PKreplicas}
\end{figure}

\begin{figure} 
\centering
\includegraphics[width=\hsize]{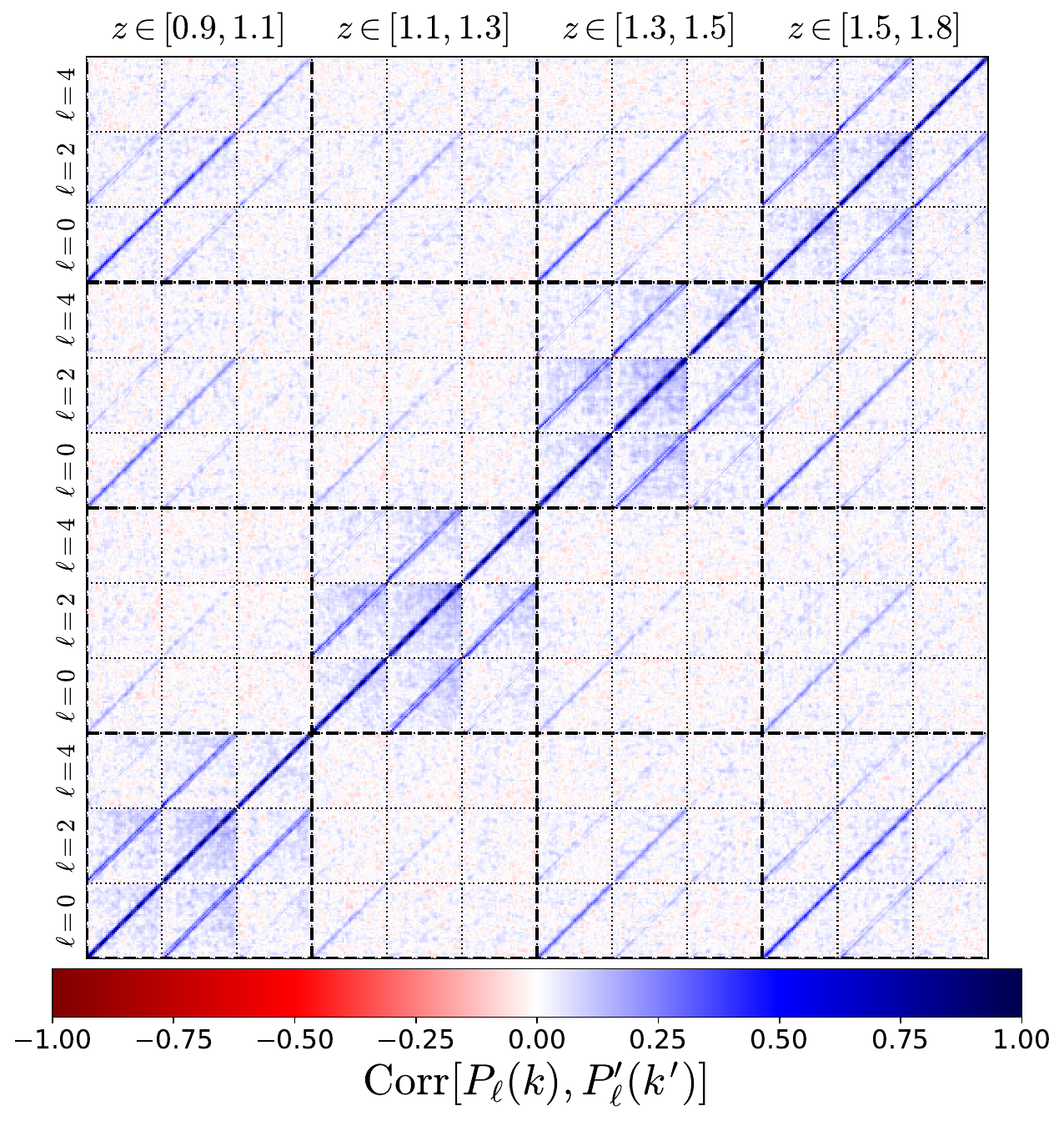}
\caption{Correlation matrix of the monopole of the power spectrum of dark matter haloes
  with $M_{\rm h}>10^{12}$\,\hmpc, measured on the four redshift bins with edges in
  $[0.9,1.1,1.3,1.5,1.8]$, for the Geppetto sets. The spurious cross-correlations due to
  replications are evident.}
\label{fig:covreplicas}
\end{figure}

\subsection{The problem of replications}
\label{sec:replicas}

The starting point of any simulation is a realisation of the linear density field on a
periodic box of length $L_{\rm s}$, that samples the Fourier space modes on a regular grid
with a cell size equal to the box fundamental mode $k_{\rm fs} = 2\pi/L_{\rm s}$. While,
as illustrated in Fig.~\ref{fig:replicas}, the survey volume is fully contained in the
EuclidLargeBox volume, in the Geppetto case the box is tiled several times to cover the
light-cone. As pointed out in \cite{Howlett2015}, when the measurement of the power
spectrum is performed, a redshift bin is cut from the cone volume and immersed in a box of
size $L_{\rm m}$ that is at least twice as large as the largest distance (in the three
dimensions) among galaxies in the catalogue. Then the resulting density is
Fourier-transformed to measure the power spectrum. The natural sampling of this
measurement is on a Fourier-space grid of cell size $k_{\rm fm} = 2\pi/L_{\rm m}$, smaller
than $k_{\rm fs}$ and not necessarily in some harmonic relation with it. This implies an
oversampling of the original Fourier space, that becomes pretty evident when the survey
volume covers several replications of the box.

To highlight this effect, we computed the power spectrum of the dark matter haloes of the
Geppetto light-cones, with masses $>10^{12}$\,\hmsun. In order to correct for the small
difference in halo mass function between the first 600 light-cones and the remaining ones
(Fig~\ref{fig:haloMF}), we set the lower limits in halo mass at $10^{12}$\,\hmsun,
corresponding to 66 particles, in the first 600 runs and to $0.99 \times 10^{12}$\,\hmsun,
corresponding to 65 particles, in the other runs. Although the difference corresponds to
only 1 particle, it is sufficient to ensure a consistent amplitude of the halo power
spectrum. The power spectrum was computed using the \Euclid ground-segment code (Euclid
Collaboration: Salvalaggio et al., in prep.) that implements a Yamamoto--Bianchi estimator
\citep{Yamamoto2006,Bianchi2015}. We produced catalogues of DM haloes in redshift space,
in the redshift range $z\in[0.9,1.1]$, and immersed them in boxes of side $L_{\rm
  m}=7.2$\,\hgpc. The resulting power spectrum is shown in Fig.~\ref{fig:PKreplicas},
where we show averages for 100 mocks performed with V4 and with cone axis aligned with the
main diagonal (orange lines) and 100 mocks performed with V5 with random direction of the
cone axis (blue lines) to mitigate the replication problem. The figure also reports the
fundamental modes of the simulation box $k_{\rm fs}$ and of the measurement box $k_{\rm
  fm}$. The beating due to the oversampling of the Fourier space is very evident
especially in the orange lines, and the randomisation of the cone direction mitigates but
does not remove the oscillations (averaging over more realisations further dampens these
oscillations without fully removing them).

One may argue that, more than being a problem of the data set, this is a problem of the
measurement. By producing periodic boxes of 1.2\,\hgpc we are sampling the density field
with a hard limit on scales: scales larger than the box are simply not sampled. While
replications are a way to produce larger volumes with a continuous density field, one
cannot assume that they are adding information. So if the natural Fourier grid to measure
the power spectrum of the periodic box is $k_{\rm fs}$, one should use this same sampling
for the power spectrum in the light-cone. Figure~\ref{fig:PKreplicas} shows the same
measurements resampled over 6 times the fundamental mode, $k_{\rm bin} = 6\, k_{\rm fm} =
k_{\rm fs}$. It is evident that the beating disappears once the sampling of Fourier space
respects the properties of the original box. However the two measurements show
percent-level differences at large scales, that vanish toward the first baryonic acoustic
oscillation at $k\sim 0.07$\,\kmpc, that may in part be due to the different code
versions. These percent-level differences may artificially increase the covariance in a
scale range where sample variance is already important, but they are unlikely to be a
significant issue for a cosmological analysis.

Another effect of replications is to create cross-correlations of different redshift bins.
We show in Fig.~\ref{fig:covreplicas} the correlation matrix (the covariance matrix
normalised to its diagonal) of the power spectrum monopole of dark matter haloes over four
redshift bins with bin edges in $[0.9,1.1,1.3,1.5,1.8]$, obtained from the whole set of
3500 measurement. The cross-correlation of different redshift bins is expected to be
non-zero due to lensing \citep[e.g.][]{Lepori-EP19} or to systematic effects
\citep{Monaco2019}, but none of these effects is present in the mocks, so these
cross-correlations arise due to the replications, that make the density field of different
redshift bins not independent. While this is a limitation, non-null cross-covariances may
be needed only for very specific cases, and setting them to zero is a convenient way to
solve the problem in practice.

\subsection{EuclidLargeBox simulations}
\label{sec:EuclidLargeBox}

The second set of $N_{\rm real}=1000$ simulations, called EuclidLargeBox, was designed to
have a more limited set of much larger and more expensive simulations, not affected by the
replications problem. They are based on a $L_{\rm s}=3.38$\,\hgpc box sampled with $N_{\rm
  part}=6144^3$ particles. With these choices, the particle mass is $1.48\times
10^{10}$\,\hmsun, very similar to that of Geppetto simulations, and the smallest output
halo (with at least 10 particles) has a mass of $M_{\rm h}=1.48\times10^{11}$\,\hmsun.
This box is marginally smaller than that (3.6\,\hgpc) of the Flagship simulation, that is
however sampled with $16\,000^3$ particles, reaching a particle mass of $10^9$\,\hmsun.
The halo mass functions of the EuclidLargeBox set, shown in Fig.~\ref{fig:haloMF}, are
consistent to within 1\% with the ones from the Geppetto mocks. These simulations produced
much larger light-cones, covering half of the sky and starting at redshift $z=4$. Because
the \Euclid survey will never go deep in the zone of avoidance of the Milky Way, sampling
Galactic latitudes always with $|b|>20$, light-cones were produced with a sky footprint of
a circle with radius of $\theta=70^\circ$, thus reducing storage needs by 34\% with
respect to a really half-sky output. With the same aim of reducing storage, output in
periodic boxes was limited to redshifts 0 and 1. This produced 215 GB per run, 210 TB in
total, of which 80 GB per run were taken by the light-cones.

To handle the memory overhead of the fragmentation and its variations over many
realisations, it was necessary to allocate 250 bytes per particle, for a total memory
requirement of 52 TB. After a few experiments, we decided to run the code on 24 fat nodes
of the machine Galileo100 at CINECA, equipped with 3 TB each and 48 cores per node. Each run
took on average of 3 hours 45 minutes, for a computational cost of 4276 core-h per run,
amounting to a total of about 4.3 million core-h.

Due to both larger redshift range and wider sky area, the volume sampled by the light-cone
grows from 12.72\,\cgpc to 163.86\,\cgpc, that is 4.24 times larger than the simulated
box, whose volume is 38.61\,\cgpc. Indeed, the light-cone is covered by tiling the box 23
times, including tiles that have very little overlap in volume. However, much of the
volume increase is due to the higher redshift reached by the light-cone, that is relevant
for adding interlopers to the spectroscopic sample due to catastrophic redshift errors. As
shown in Fig.~\ref{fig:replicas}, the comoving volume of a survey from $z=0.9$ to $1.8$
and with a {\tdeg} radius sky footprint is contained in one simulated box. We will use the
{\tdeg} radius footprint for most of the preparation papers, so the measurements of the
EuclidLargeMocks galaxy catalogues produced from the EuclidLargeBox simulations, that are
described below, are expected to be free of the problem of replications.

The volume sampled by the Geppetto and EuclidLargeBox sets is respectively 6047\,\cgpc and
38\,614\,\cgpc, for a total of 44\,661\,\cgpc, while the light-cones have volumes of
44\,529\,\cgpc and 163\,862\,\cgpc. These can be compared to the 3400\,\cgpc of our
visible Universe. All these numbers are reported in Table~\ref{table:simulations},
together with the same numbers for two other sets of simulations produced with {\pin} and
presented in previous papers, namely the 10\,000 Minerva-like simulations used in
\cite{Oddo2020} and the 1000 NewClusterMock simulations aimed at describing galaxy
clusters and presented in \cite{Fumagalli21}. The simulation sets presented here go down
by a factor of $\sim$\,30 in halo mass, necessary to describe the spectroscopic sample
that \Euclid will observe.

This archive of data sets can be compared with other simulation sets available in the
literatures, as the remarkable
Quijote\footnote{\url{https://quijote-simulations.readthedocs.io}} project
\citep{Quijote}, that samples a volume of 82\,000\,\cgpc in boxes of 4\,\hgpc of side.
While the volume covered by Geppetto and EuclidLargeBox together is only half of it, the
total volume produced with {\pin} for \Euclid studies, including the low-resolution boxes,
is even larger than this ground-breaking figure; of course, our sets do not (yet) cover
that rich variety of cosmologies. Our data set is unique in the box size, while still
achieving a suitable mass resolution to sample galaxies, and in the number of light-cones,
that presently provide the largest data set of dark matter haloes in the light-cone ever
produced and shared with the cosmological community.

\section{From haloes to galaxies}
\label{sec:hod}

In this section we describe the procedure that we implemented to obtain \Euclid
spectroscopic skies from our collection of halo light-cones, thus bridging the gap from
unobservable dark matter haloes to visible galaxies.

The reference data set for \Euclid preparation papers is the Flagship simulation and mock
catalogue described in the Flagship paper. We aim at creating a set of galaxy catalogues
that have the same mean number density and 2-point clustering as a spectroscopic sample
extracted from the Flagship mock, for a broad range of selections. All the galaxy
catalogues that we produce are contained in a {\tdeg} radius footprint, as the Geppetto
light-cones; the EuclidLargeBox footprint is much larger, but to have a consistent
approach in the test of the galaxy power spectrum and 2-point correlation function we
decided to use, across several \Euclid preparation papers, this footprint; we thus call
EuclidLargeMocks the galaxy catalogues extracted from the EuclidLargeBox set of
simulations using a {\tdeg} radius footprint. This footprint can be fully immersed in the
Flagship octant, all comparisons with Flagship will be done consistently using this
footprint.

\begin{figure} 
\centering
\includegraphics[width=\hsize]{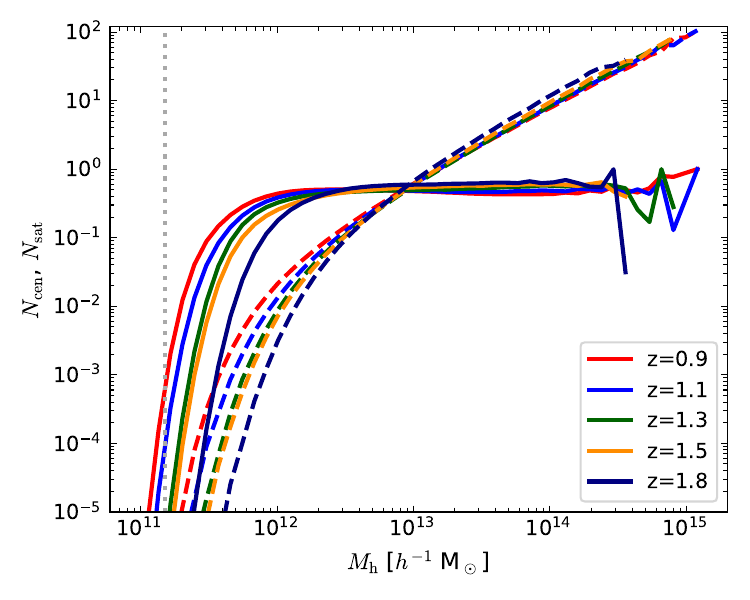}
\caption{HOD curves $N_{\rm cen}(M_{\rm h},z)$ and $N_{\rm sat}(M_{\rm h},z)$,
  respectively as continuous and dashed lines, measured from the Flagship mock catalogue
  as described in the main text. The five colours give the curves at five redshifts, as
  specified in the legend. The vertical dotted line gives the minimum halo mass in the
  \pin mocks.}
\label{fig:hod}
 \end{figure}

The steps that lead from the Flagship $N$-body simulation to a set of galaxies with
measurable {\ha} flux are thoroughly described in the Flagship paper. The procedure is
based on an HOD conceived and calibrated to produce a very large range of galaxy
properties, with the flux of galaxy emission lines coming at the end of the process. This
makes it impossible to analytically infer HOD curves for galaxies selected in {\ha} line
flux \fha. The number density of galaxies with \fha larger than a threshold (of order of
$f_0$ defined in Sect.~\ref{sec:intro}) is close to model 3 of \cite{Pozzetti2016}. When
plotted as a function of redshift, this number density shows little blips due to an issue
with the interpolation of internal galaxy extinction; these blips have no appreciable
practical consequences but will be sometimes visible in the figures that we will show
below.

It is not convenient to directly implement such an assignment scheme to populate thousands
of Flagship-like simulations; we then bypass this whole procedure by measuring the HOD
functions directly from the Flagship catalogue. The algorithm, very similar to that used
in the perturbation theory challenge \citep[][where it was applied to Flagship haloes in
  the periodic box]{EP-Pezzotta}, runs as follows. We first set a selection criterion for
the galaxy catalogue, based on a measured galaxy property, in this case the {\ha} line
flux: $\fha>f_{\rm lim}$. We repeat the selection using a very fine grid of limiting flux
values $f_{\rm lim}$, starting from $0.5 f_0=10^{-16}$ \flux, below which we assume that a
reliable measure of redshift is very unlikely. For each flux cut we measure the two HOD
curves relative to central and satellite galaxies in a grid of halo mass and redshift
(without redshift-space distortions):

\begin{eqnarray} N_{\rm cen}(M_{\rm h}, z\, |\, \fha\ge f_{\rm lim})&=&\frac{\rm
    \#\ of\ central\ galaxies}{\rm \#\ of\ haloes}\, ,
\label{eq:Ncen}\\
 N_{\rm sat}(M_{\rm h}, z\,|\,  \fha\ge f_{\rm lim})&=&\frac{\rm
   \#\ of\ satellite\ galaxies}{\rm \#\ of\ haloes}\, ,
\label{eq:Nsat}
\end{eqnarray}

\noindent
where the numbers are relative to that specific bin and flux cut. Here $N_{\rm cen}$
should be interpreted as the probability that a halo of given mass and at a given redshift
hosts a central galaxy that satisfies our selection criterion, $N_{\rm sat}$ as the
average number of satellites in that same halo. We choose a redshift bin $\delta z=0.01$
from 0 to 3 (where the Flagship mock ends) and a mass bin of $\delta \logten (M_{\rm
  h}/M_\odot) = 0.083$; to limit noise from sample variance we smooth these curves in
redshift with a Gaussian kernel of width of five bins. The resulting curves are presented
in Fig.~\ref{fig:hod}, as a function of $M_{\rm h}$, for $\fha>f_0$ and for several
redshifts from 0.9 to 1.8. While $N_{\rm sat}$ can overshoot unity, $N_{\rm cen}$ is
constrained to be $\le1$, and its values flatten at large masses at $\sim$\,0.4. The plots
also report the mass of the smallest halo in our Geppetto and EuclidLargeBox simulations
as a vertical dotted line, to show that the number of {\ha}-emitting galaxies hosted in
dark matter haloes below our mass resolution is expected to be negligible, at least
according to this HOD.

The same algorithm can be used to work out HOD curves for any other observational
selection, like a limit in the \Euclid magnitude $\HE<24$ that defines the photometric
sample for which a measurement of the spectrum is attempted. However, the mass resolution
of our simulation sets has been chosen to resolve the haloes that contain galaxies of the
\Euclid spectroscopic sample, but galaxies of the photometric sample are hosted by smaller
haloes and so a photometric mock catalogue would be incomplete, especially at low
redshift.

\begin{figure}
\centering
\includegraphics[width=\hsize]{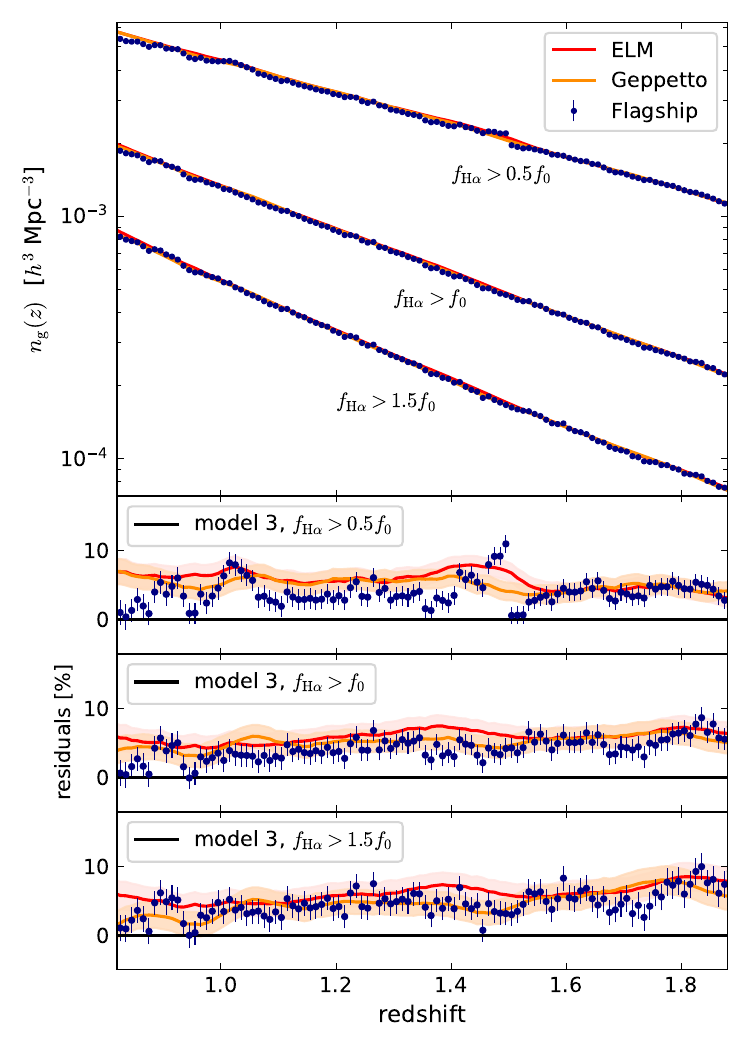}
\caption{Comoving number density $n_{\rm g}$, in \dens, of galaxies as a function of
  redshift for Geppetto (orange), and EuclidLargeMocks (red) catalogues, compared with the
  Flagship catalogue (dark blue points). For the sets of {\pin} catalogues we give the
  average number density as a continuous line and its standard deviation as a shaded area,
  while Flagship values, reported as circles, are assigned the EuclidLargeMocks variance,
  reported as an errorbar. We show in the top panel results for three limiting fluxes,
  $\fha>0.5 f_0$, $f_0$, and $1.5f_0$. The other panels give, for each limiting flux, the
  residuals ($n_{\rm g}/n_{\rm model\,3}-1$) in percent of simulated number densities with
  respect to model 3 from \cite{Pozzetti2016}. The black lines denote the zero value of
  the residuals.}
\label{fig:dndz}
\end{figure}

\begin{figure} 
\centering
\includegraphics[width=\hsize]{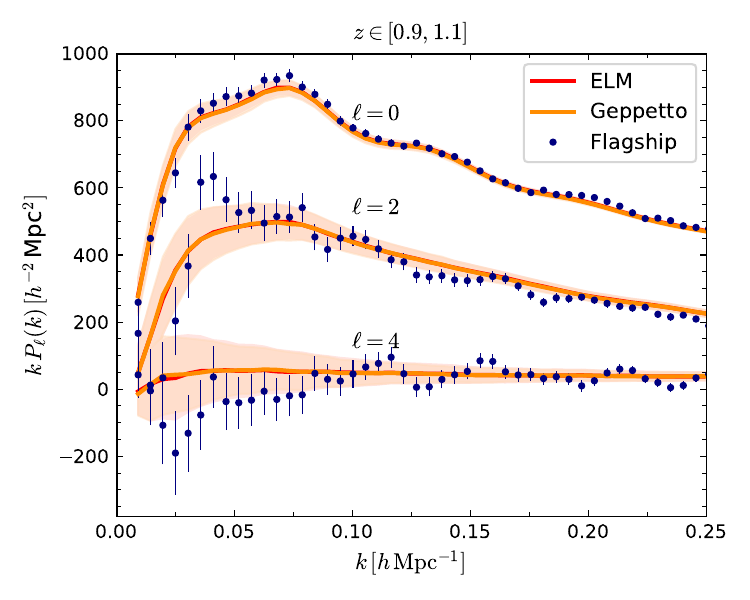}
\caption{Power spectrum multipoles of galaxies with $\fha>f_0$, for Flagship (blue
  points), the first 1000 Geppetto mocks (orange lines and shaded areas), and the 1000
  EuclidLargeMocks (ELM, red lines and shaded areas). We show here the first redshift bin,
  $z\in[0.9,1.1]$. For the \pin sets we report the average measurement as a line and its
  variance as a shaded area; Geppetto and ELM lines are very similar and hard to
  distinguish. The Flagship points are assigned an errorbar equal to the standard
  deviation of the EuclidLargeMocks.}
\label{fig:PK_elmvsgep}
\end{figure}

To each halo we associate a single central galaxy, with a probability $N_{\rm cen}$, and a
number of satellite galaxies drawn from a Poissonian distribution with mean $N_{\rm sat}$.
Central galaxies inherit the halo position and velocity, while satellite galaxies are
distributed following a Navarro--Frenk--White profile \citep[NFW,][]{Navarro1996}, with
concentration given by the \cite{Diemer2019} relation and velocities around the halo
centre of mass with magnitude

\be
V_{\rm sat}(r) = f_{\rm v}\, \sqrt{\frac{GM_{\rm h}(<r)}{r}}\, ,
\label{eq:vsat}\ee

\noindent
where $f_{\rm v}$ is a calibration constant and $M_{\rm h}(<r)$ is the halo mass within
distance from halo centre $r$, given by the NFW profile. Orientations of satellite
velocities are randomly drawn from a sphere. The calibration constant was added to recover
the velocity dispersion of Flagship satellites, that is computed with a more sophisticated
scheme aimed at reproducing in detail the velocity dispersion of galaxies in cluster. Its
value after calibration is $f_{\rm v}=0.7$.

{\pin} and $N$-body halo masses are not expected to be equivalent, both because {\pin}'s
claimed accuracy is $\sim$\,5\% in the halo mass function and because it has been
calibrated to reproduce haloes found with the friends-of-friends algorithm, not those
found with the \texttt{Rockstar} halo finder \citep{Behroozi2013} used in the Flagship
simulation (see \citealt{Castro-EP24} for a discussion of the differences among halo
finders). Moreover, {\pin} is able to reproduce the linear halo bias to within a few
percent \citep{Paranjape2013,Munari2017}, and this difference in clustering amplitude can
be absorbed by halo calibration. This was done in \cite{Oddo2020}, where an optimal
reproduction of the halo bispectrum and its covariance was achieved by calibrating the
mass cuts of simulated and {\pin} haloes so as to have the same amplitude of the power
spectrum. To apply this HOD to our halo light-cones, we calibrate halo masses with a
clustering matching (CM) procedure, described in detail in the Appendix, and different
from the abundance matching (AM) procedure that would guarantee consistency in the galaxy
number density. As discussed in the Appendix, this match (obtained with a simple
procedure) guarantees the same clustering amplitude for the halos containing the bulk of
the {\ha} galaxies and is adequate for the present purposes, but it should be made more
sophisticated to better account for the mass dependence of the accuracy of halo bias. Once
the clustering level is matched, matching of number density is achieved by multiplying the
HOD curves $N_{\rm cen}$ and $N_{\rm sat}$ by the ratio of a target number density and the
measured one. Indeed, as long as the largest value of $N_{\rm cen}$ remains below unity,
these curves can be shifted up or down with no effect on the clustering amplitude (apart
the obviously different shot noise level), as the changing amplitude leads to a different
sampling of the parent halo distribution, keeping the ratios of haloes of different mass
constant.

\begin{figure*} 
\centering
\includegraphics[width=1.0\hsize]{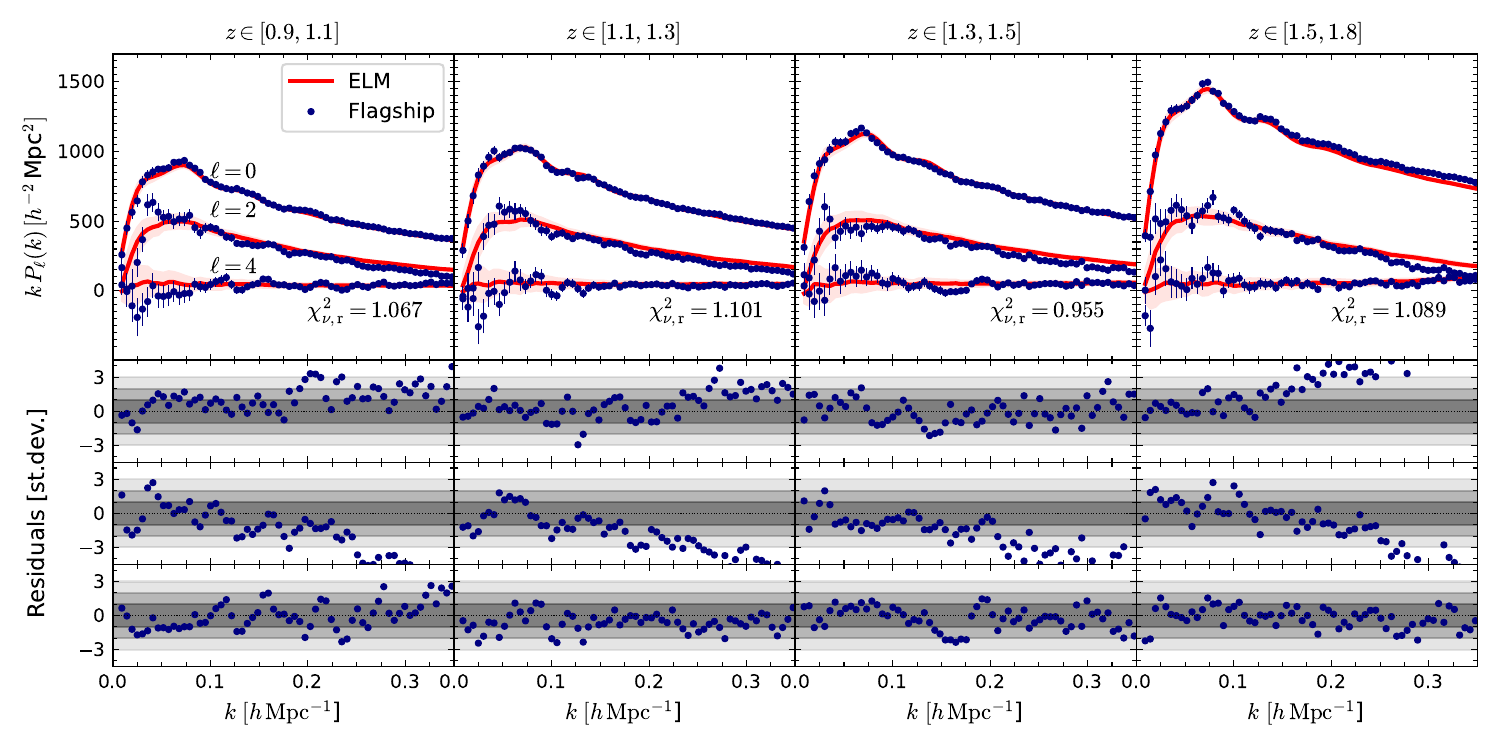}
\caption{Power spectrum multipoles of galaxies with $\fha>f_0$, for Flagship (blue
  points), and EuclidLargeMocks (red lines), in the four redshift bins, as indicated at
  the top of each column of panels. Lower panels show the residuals of the Flagship
  measurements (from top: monopole, quadrupole, hexadecapole) with respect to the average
  EuclidLargeMocks in units of the standard deviation of the latter.}
\label{fig:PKall}
\end{figure*}

The number density of the Flagship catalogue was calibrated to reproduce model 3 of
\cite{Pozzetti2016}, achieving an accuracy of roughly 5\%, larger than the expected sample
variance though not larger than the observational uncertainty. As a target number density
we then used that of the Flagship catalogue. We fitted the ratio of the number density
obtained with CM, averaged over 100 mocks, and the Flagship density with a 4th-order
polynomial, and used this fitted ratio to modulate the two HOD curves as a function of
redshift. We give the resulting number density $n_{\rm g}$ in Fig.~\ref{fig:dndz},
compared with that of the Flagship mock on the same {\tdeg} radius footprint and with
model 3, for three flux limits, $\fha > 0.5f_0$, $f_0$, and $1.5f_0$, amounting to $1$,
$2$ and $3\times10^{-16}$ {\flux}. For the Geppetto and EuclidLargeMocks sets we report
the standard deviation of the number density as a shaded area; the same standard deviation
(from the EuclidLargeMocks) is assigned to the Flagship and shown as an errorbar. The
three lower panels report the difference of the number densities with respect to model 3
predictions (not reported in the above panel) for the three flux limits. The Flagship mock
fluctuates around the Geppetto and EuclidLargeMocks average values in a way that is
compatible with the expected standard deviation, with some difference at the lowest
redshift; at low flux the little blips in the Flagship number density, at $z=1$ and $1.5$,
is due to the issue connected to the interpolation of internal extinction mentioned in
Sect.~\ref{sec:hod}.

\begin{figure*} 
\centering
\includegraphics[width=\hsize]{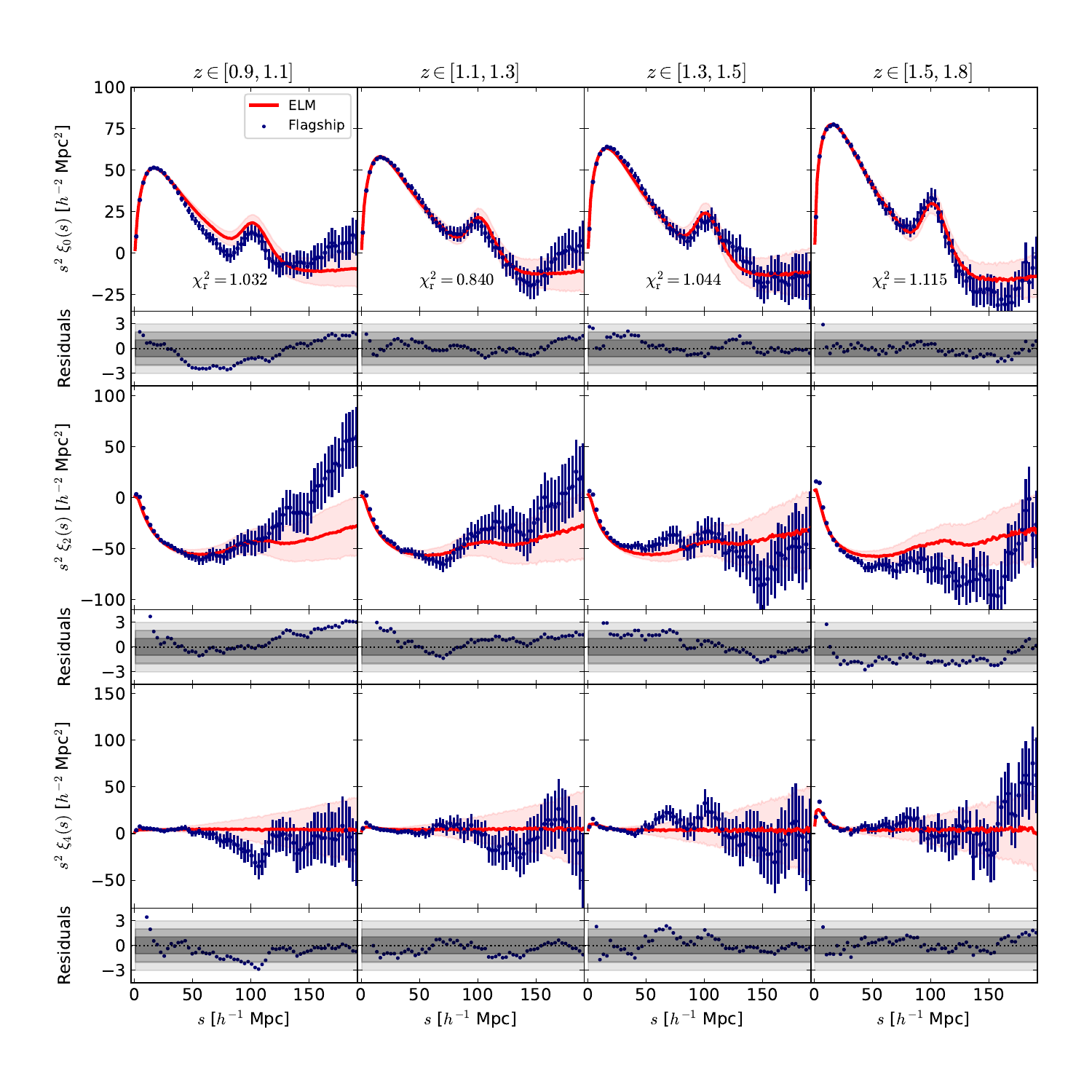}
\caption{Two-point correlation function multipoles of galaxies with $\fha>f_0$, for
  Flagship (blue points), and EuclidLargeMocks (red lines), in the four redshift bins as
  indicated at the top of each column of panels. Lower panels show the residuals of the
  Flagship measurements with respect to the average EuclidLargeMocks in units of the
  standard deviation of the latter.}
\label{fig:2PCFall}
\end{figure*}

We conclude this section by arguing that the small differences in number density and halo
clustering between $N$-body and {\pin} haloes found in Sect.~\ref{sec:simulations} are not
relevant in practice as they can be calibrated away. The aim of these simulation sets is
to provide the covariance of clustering measurements of an observed galaxy sample, and
this aim is achieved by calibrating the HOD (or in general the algorithm used to populate
haloes with galaxies) to reproduce its properties. Suppose that one calibrates an HOD on
the galaxy number density and 2-point clustering amplitude, and performs this calibration
independently using an $N$-body and a {\pin} halo light-cone. The small differences in
halo properties will result in slightly different HOD parameters, with virtually identical
results on the calibrated quantities. Such small differences will have negligible impact
on the clustering covariance, whose leading terms are related to the amplitude of the
power spectrum, that is matched to the data. Moreover, the accuracy requirement on the
covariance is weaker than the one on the average measurement, so percent-level effects
will likely be negligible. At the same time, given that we are not predictive in how a
specific galaxy sample populates DM haloes, we can consider the HOD as a nuisance for
cosmological parameter inference; as long as we are able to effectively marginalise over
nuisance parameters, the details of the HOD will be immaterial to the final results.

\section{Validation}
\label{sec:validation}

In this section we present the validation tests that we have performed to demonstrate that
the mocks can faithfully represent the covariance of the Flagship galaxy mock. Most effort
is devoted to test the EuclidLargeMocks set, that is not affected by the issue of
replications (Sect.~\ref{sec:replicas}), but given the very consistent results of the code
in the two configurations one can safely extend the validity of the tests to the Geppetto
set, at least on the scales well sampled by the simulation box of 1.2\,\hgpc. In all cases
the Flagship is projected on a {\tdeg} radius circle to have the same footprint as the
other mock catalogues.

\begin{figure*} 
\centering{
\includegraphics[width=0.45\hsize]{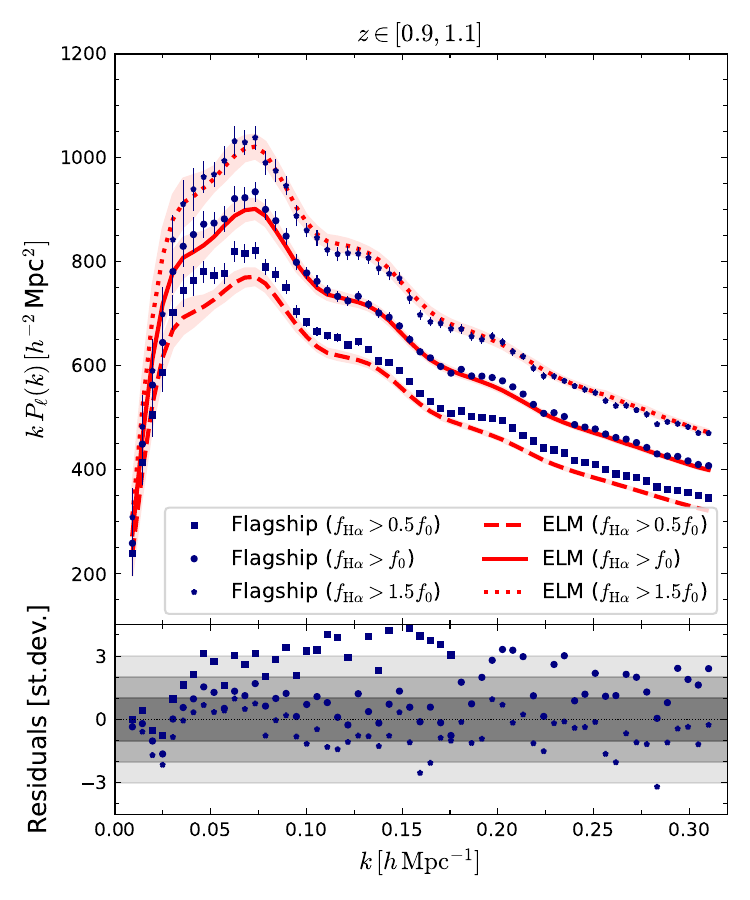}
\includegraphics[width=0.45\hsize]{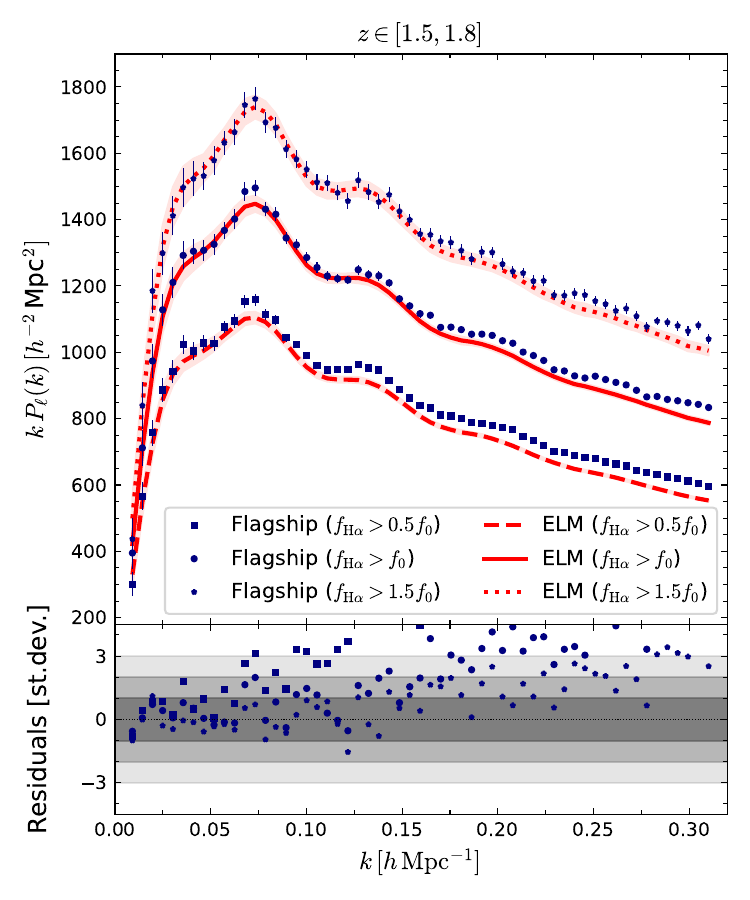}
}
\caption{Power spectrum monopole of galaxies in the Flagship mock and in the
  EuclidLargeMocks, for $\fha>0.5f_0$, $f_0$, and $1.5f_0$, and for $z\in [0.9,1.1]$ (left
  panel) and $z\in[1.5,1.8]$. Lower panels show the residuals of the Flagship measurements
  for the three flux limits with respect to the average EuclidLargeMocks in units of the
  standard deviation of the latter.}
\label{fig:PKflux}
\end{figure*}

\begin{figure*} 
\centering{
\includegraphics[width=0.49\hsize]{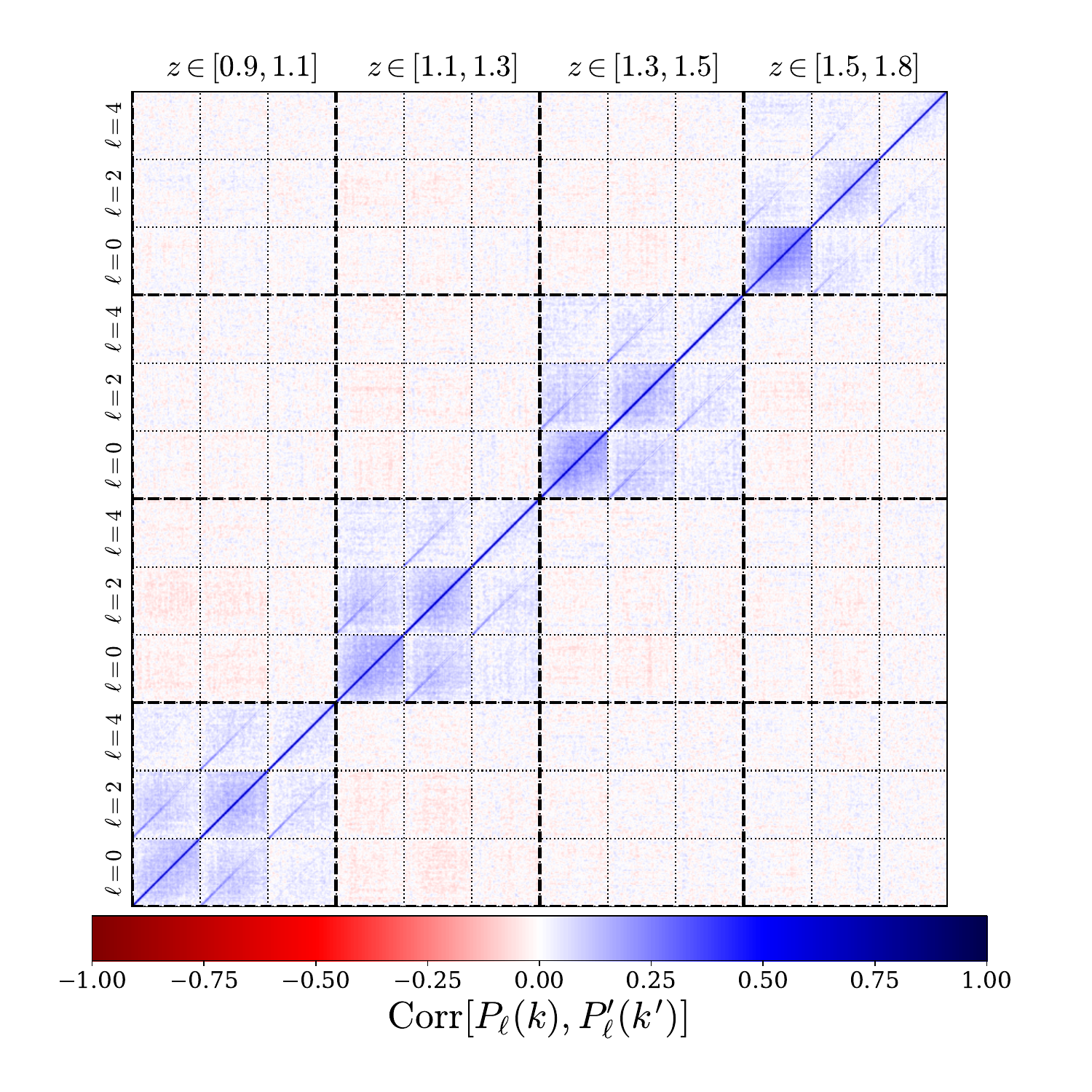}
\includegraphics[width=0.49\hsize]{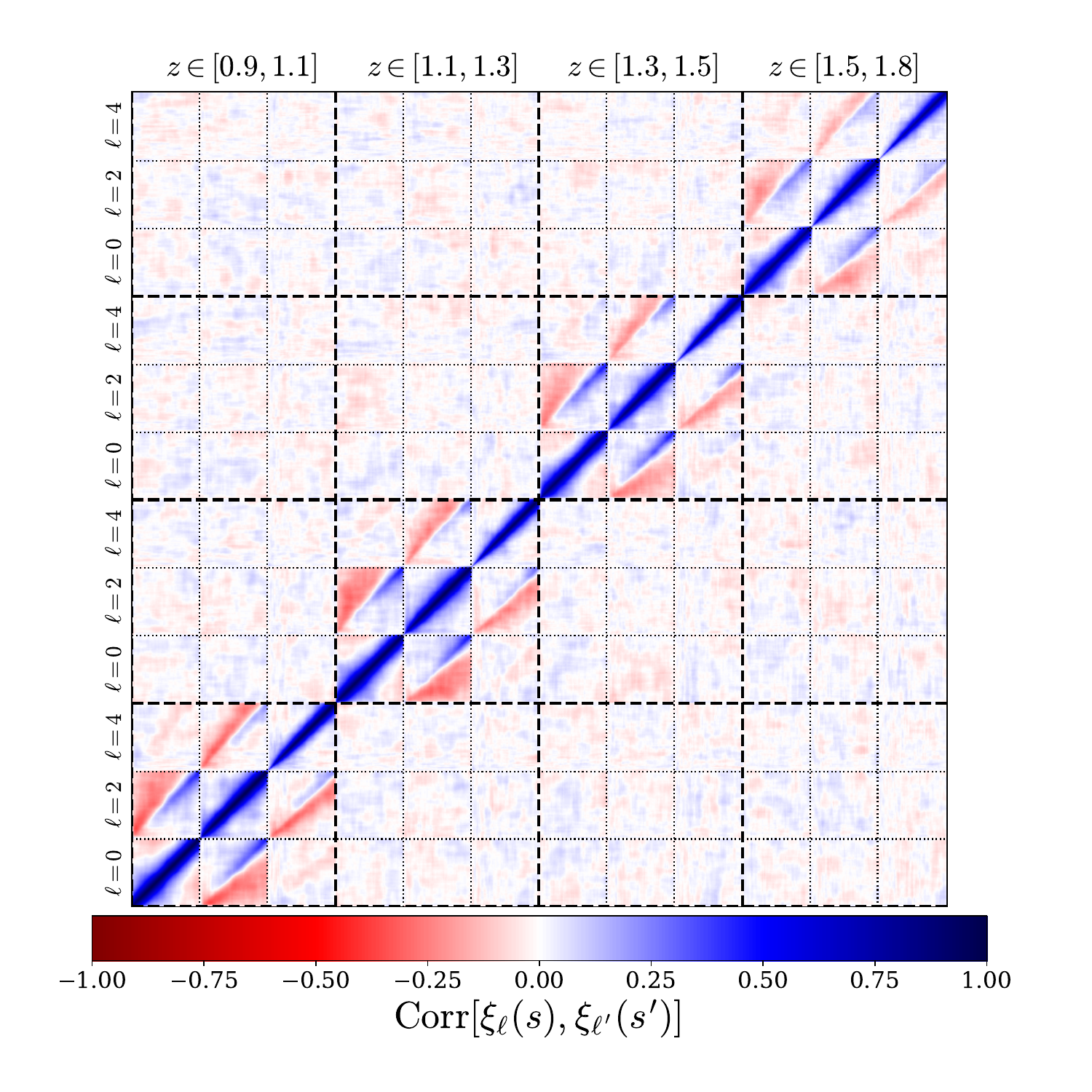}}
\caption{Correlation matrices of the first three even multipoles of the power spectrum
  (left) and of the 2-point correlation function (right), for all the redshift bins.}
\label{fig:covariances}
\end{figure*}

\begin{figure*}
\centering
\includegraphics[width=\hsize]{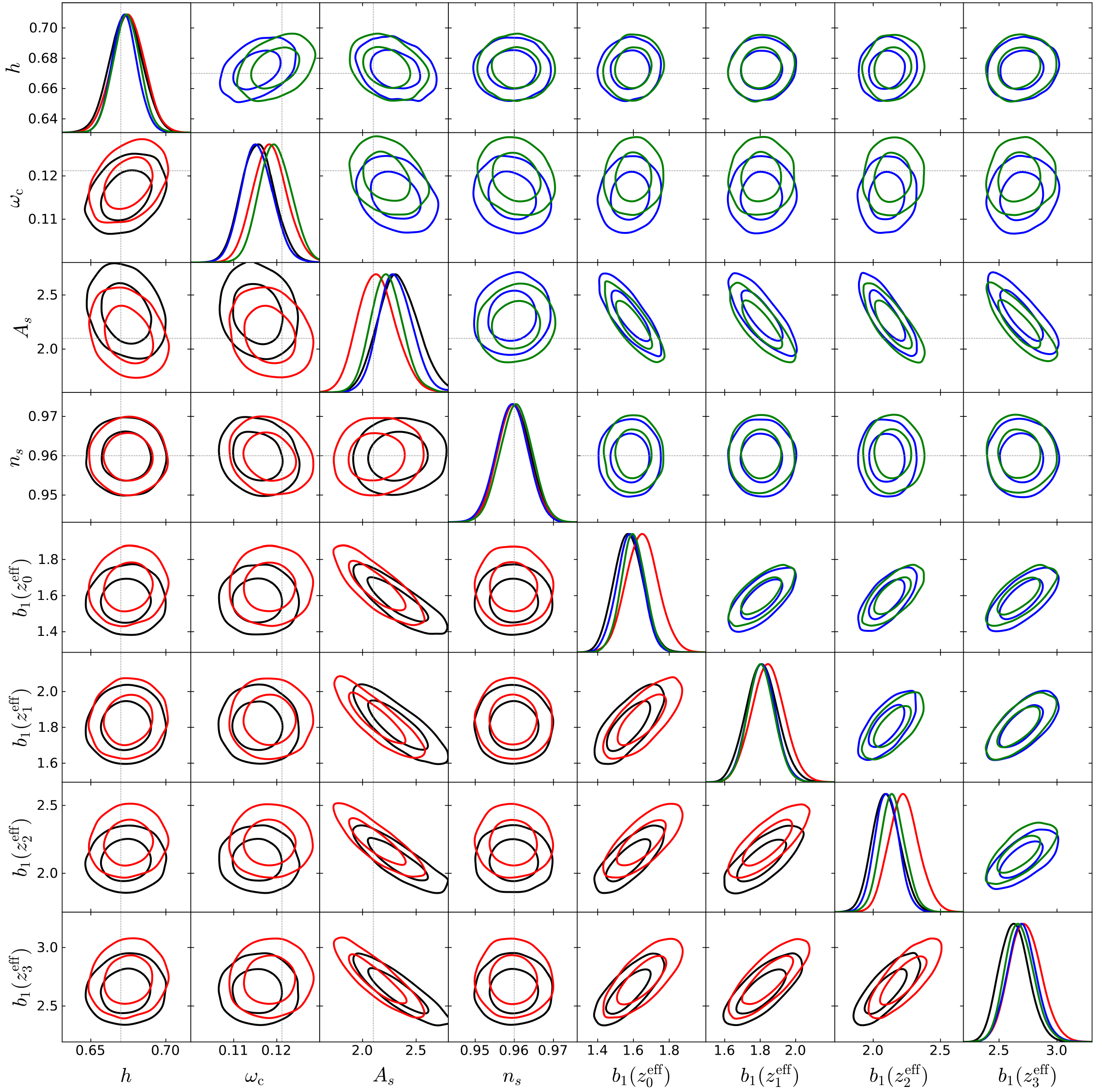}
\caption{Posteriors of parameter inference obtained by fitting the Flagship (black and
  blue lines) and one of the EuclidLargeMocks (red and green lines) with cuts at
  $k<0.2$\,\kmpc (lower-left corner with red and black lines) and $k<0.3$\,\kmpc
  (upper-right corner with blue and green lines).}
\label{fig:definitive_test}
\end{figure*}

We first test the results for the galaxy power spectrum at the nominal flux limit $f_0$,
for both the Geppetto and the EuclidLargeMocks sets. We use all the 1000 EuclidLargeMocks
and the first 1000 Geppetto mocks, the binning of the power spectrum is at 6 fundamental
modes to suppress the beating of Geppetto measurements due to replications. We show in
Fig.~\ref{fig:PK_elmvsgep} the first three even multipoles of the first redshift bin,
$z\in[0.9,1.1]$. The power spectra of Geppetto and EuclidLargeMocks are so similar that it
is hard to distinguish them in the figure, both in the mean and in the standard deviation;
in particular, a (very) close analysis of the figure shows that the standard deviation of
the Geppetto mocks is some $\sim$\,5\% smaller than that of the EuclidLargeMocks, and this
is broadly consistent with the missing super-sample covariance in the smaller simulated
volume, while adding together simulations performed with different code versions does not
seem to contribute. Both average power spectra reproduce very closely the measurement of
the Flagship mock in the same footprint; for this summary statistics Flagship is
consistent with being a realisation drawn from the set, the largest discrepancy being a
flare in the Flagship quadrupole on large scales that is however consistent with being a
statistical fluctuation.

To give a more comprehensive view of the results, Figs.~\ref{fig:PKall} and
\ref{fig:2PCFall} show the power spectrum and 2-point correlation function multipoles of
the galaxy catalogue cut at the fiducial flux limit, $\fha>f_0$, for the four redshift
bins bounded by $[0.9,1.1,1.3,1.5,1.8]$, for Flagship and for the 1000 EuclidLargeMocks.
Each column of panels reports the value of the reduced chi-squared, $\chi^2_{\rm r}$
relative to that redshift bin, computed for $k<0.2$\,\kmpc (power spectrum) and $r$
between 50 and 150 \hmpc (two-point correlation function), using the three multipoles of
the Flagship mock as data vector and the EuclidLargeMocks as model and covariance (see
Fig~\ref{fig:covariances} below). In all cases the agreement is excellent: for the power
spectrum, and in the wavelength range $k<0.2$\,\kmpc typically used in standard inference
of cosmological parameters, the Flagship measurement is consistent with being drawn from a
realisation of the EuclidLargeMocks, while at higher wave-numbers the EuclidLargeMocks
tend to overestimate its quadrupole, and underestimate the monopole in the last redshift
bin by 7\%. As mentioned in the Appendix, we calibrated the CM relation by requiring an
accuracy better than 10\% on the scales dominated by the one-halo term, so this difference
can be decreased by improving the HOD calibration. The overestimation of the quadrupole
was already noticed by \cite{Munari2017} and, not only for the \pin code, by
\cite{Blot2019}, and will need a specific mitigation strategy if the scale cut is going to
be more aggressive than $k<0.3$\,\kmpc. These differences are less visible in the 2-point
correlation function (Fig.~\ref{fig:2PCFall}), where the (much more correlated)
measurements of the Flagship mock are always consistent with being drawn from the
distribution of the EuclidLargeMocks. Here the small apparent shift in the
baryonic-acoustic-oscillation (BAO) peak visible in the monopole is consistent with sample
variance; this is evident from the low $\chi^2_{\rm r}$ values, but we also checked that
fitting the BAO peak for the average measurement and for the Flagship mock provides very
consistent results (Euclid Collaboration: Sarpa et al., in prep.). As noticed above, in
both estimators the Flagship measurements show large and sometimes significant
($\sim$\,3$\sigma$) flares of the quadrupole on large scales.

The accuracy with which luminosity-dependent galaxy bias is reproduced is shown in
Fig.~\ref{fig:PKflux}. We report here the monopole of the power spectrum for flux cuts
$\fha>0.5f_0$, $f_0$, and $1.5f_0$, in the first and last redshift bin ($z\in[0.9,1.1]$
and $z\in[1.5,1.8]$). Luminosity-dependent bias is well reproduced, although at the lowest
flux we notice an underestimation of the monopole that would require a more sophisticated
implementation of CM to be removed (see the discussion in the Appendix); 
the number of galaxies detected in the actual survey
at such low fluxes is expected to be small, so this discrepancy is unlikely to be a
significant problem.

These measurements allow us to construct a numerical covariance matrix that takes into
account in full detail the geometry of the survey. Figure~\ref{fig:covariances} shows the
correlation matrices of all the multipoles of the power spectrum for all the redshift
bins, on the left for the power spectrum and on the right for the 2-point correlation
function. In this case the correlation of different redshift bins is very low, as
expected, showing that the problem of replications is absent in the EuclidLargeMocks.

We use these covariances in one example of inference, similar to what is done in Euclid
Collaboration: P. Monaco et al. (in prep.). We fit the Flagship power spectrum multipoles
across four redshift bins using a model based on the effective field theory of large-scale
structure \citep[see][for a recent review]{cabass_snowmass_2022}. Model and likelihood
evaluations are performed using the Gaussian process emulator \texttt{Comet}
\citep{eggemeier_comet_2022}, while posterior estimation is conducted with
\texttt{NAUTILUS} \citep{lange_nautilus_2023}. We fit the monopole and quadrupole of the
power spectrum, binned in $k$ with an initial bin $k_{\rm i} = 6\, k_{\rm F} \sim
5.4\times 10^{-3}\, \kmpc$ and bin size $\Delta k = 6 \,k_{\rm F} $. The maximum
wave-number $k_{\rm max }$ is varied across different fits. The model includes five
cosmological parameters: $\{ h,\omega_{\rm c},A_{\rm s}, n_{\rm s},\omega_{\rm b}\}$,
where $\omega=\Omega\, h^2$ and c stands for cold dark matter, b for baryons, while
$A_{\rm s}$ and $n_s$ are the normalisation and the spectral slope of primordial
perturbations. We adopt wide uniform priors for all parameters except for the spectral
index $n_{\rm s}$, which follows a Gaussian prior $\mathcal{N} (0.96,0.041)$, and the
baryon density parameter, which follows $\mathcal{N} (2.218\times 10^{-2},0.055 \times 10
^{-2})$. Additionally, we vary three bias parameters $\{ b_1,b_2,\gamma_{21} \}$ for each
redshift bin and analytically marginalise over the counter- and shot noise terms.

We also use one of the measurements from the EuclidLargeMocks as a data vector and rerun
the chains to get another set of parameter values. Figure~\ref{fig:definitive_test} shows
the resulting parameter posteriors (with the exception of $\omega_b$, that is
prior-dominated), where the black and red lines in the lower left triangle give results
obtained respectively with Flagship and EuclidLargeMocks using a conservative cut of
$k<0.2$\,\kmpc, while the blue and green lines in the upper-right triangle give results
obtained using a more aggressive $k<0.3$\,\kmpc cut. In all cases the contours are
consistent among themselves to within 1$\sigma$, while inferred values of cosmological
parameters are consistent with the true ones. This test demonstrates that the
EuclidLargeMocks can be safely used to do parameter inference for \Euclid galaxy
clustering.

\section{Prospects and conclusions}
\label{sec:conclusions}

We presented the largest collection of simulated dark matter haloes in the light-cone ever
produced, and a corresponding set of 3500$+$1000 galaxy mock catalogues of the \Euclid
spectroscopic sample that cover each an area of 2763 {\sqdeg}, larger than the planned
area of DR1 in \cite{Scaramella-EP1} and roughly one fifth of the final EWS. The
simulations were produced with the {\pin} code for the generation of approximate
catalogues of dark matter haloes, in a numerical effort of $\sim$\,5\,000\,000 core-h that
produced $\sim$\,300 TB of halo catalogues. The galaxy mock catalogues were obtained by
extracting an HOD from the Flagship mock catalogue described in the Flagship paper, and
calibrating it to absorb the little differences in the clustering and number density of
{\pin} and simulated haloes.

We validated these sets of galaxy mock catalogues by comparing number densities, power
spectra and 2-point correlation functions, of haloes and galaxies with different halo mass
or line flux cuts and in four redshift bins from $z=0.9$ to $1.8$, with the same
measurements of the Flagship spectroscopic catalogue. We found that the spectroscopic
sample drawn from the Flagship mock catalogue is consistent with being a realisation drawn
from one of our sets, and this is testified by the reasonable values of $\chi^2_{\rm r}$
obtained using the Flagship multipoles as a data vector and the EuclidLargeMocks for the
model and the covariance. We identified two main limitations of our mocks, in particular
the Geppetto set is affected by the relatively small box size, replicated several times to
fill the light-cone volume. This results in spurious oscillations of the galaxy power
spectrum that can be mitigated by performing the measurements in units of the fundamental
mode of the simulation box. This problem does not affect the EuclidLargeMocks, where the
survey volume is contained in the simulation box. Also, For both sets we noticed an
overestimation of the power spectrum quadrupole at $k>0.2$\,\kmpc. Conversely, we noticed
a flare of the quadrupole of the Flagship galaxies on the largest scale that is consistent
with being a statistical fluctuations, while some blips in the Flagship galaxy number
density are due to a known issue in the interpolation of internal extinction needed to
compute the {\ha} line flux.

As a test of this consistency, we verified that using the power spectrum of a single
EuclidLargeMocks or the Flagship mock for inferring parameters in a maximum likelihood
fit, where the numerical covariance is obtained from the EuclidLargeMocks, gives
consistent posteriors for all the cosmological and nuisance parameters. This remains true
even when pushing the scale cut to $k<0.3$\,\kmpc, showing that the overestimation of the
quadrupole does not bias cosmological inference. This fully demonstrates the usability of
these mock catalogues for analysing \Euclid's DR1.

These mocks were produced for \Euclid's preparation. In particular,
the EuclidLargeMocks are being extensively used in many \Euclid preparation papers,
especially in the papers of the Observational Systematics Key Project presented by Euclid
Collaboration: P. Monaco et al. (in prep.), and in the papers of the Organisation Unit for
Level-3 products, in particular for testing the estimator of the power spectrum (Euclid
Collaboration: Salvalaggio et al., in prep.). 
Because they are calibrated on the Flagship simulation and not on real data, the scope 
of the galaxy mock catalogues is limited to such preparatory work. The halo light-cones will be
reprocessed with an updated HOD to produce more realistic DR1-like catalogues, and will thus have a 
higher longevity. While nothing prevents us to use the EuclidLargeBox light-cones for the 
whole survey, for DR2 and DR3 we are planning an even more extended suite of simulations
run with a code that improves on the scalability of the fragmentation part, and possibly with a
larger simulation volume and better mass resolution.

\section{Data availability}

Simulated catalogues of dark matter haloes on the light-cone are available at
\url{https://adlibitum.oats.inaf.it/pierluigi.monaco/euclid_mocks.html}, Euclid mock
catalogues are available through the CosmoHub platform \citep{Carretero:17,Tallada:20},
\url{https://cosmohub.pic.es}.

\begin{acknowledgements}
{\AckEC} {\AckCosmoHub} 
This paper has been supported by: the Fondazione ICSC, Spoke 3 Astrophysics and Cosmos
Observations. National Recovery and Resilience Plan (Piano Nazionale di Ripresa e
Resilienza, PNRR) Project ID CN\_00000013 `Italian Research Center on High-Performance
Computing, Big Data and Quantum Computing' funded by MUR Missione 4 Componente 2
Investimento 1.4: Potenziamento strutture di ricerca e creazione di `campioni nazionali di
R\&S (M4C2-19 )' - Next Generation EU (NGEU); by the National Recovery and Resilience Plan
(NRRP), Mission 4, Component 2, Investment 1.1, Call for tender No. 1409 published on
14.9.2022 by the Italian Ministry of University and Research (MUR), funded by the European
Union – NextGenerationEU– Project Title `Space-based cosmology with Euclid: the role of
High-Performance Computing' – CUP J53D23019100001 - Grant Assignment Decree No. 962
adopted on 30/06/2023 by the Italian Ministry of Ministry of University and Research
(MUR). Computing time was obtained from CINECA ISCRA-B grant `EuMocks', from INFN and from
the Pleiadi system of INAF. \citep{Taffoni2020,Bertocco2020}
\end{acknowledgements}

\bibliography{mybiblio}

\begin{appendix}
  
\section{Calibration of {\pin} and Flagship halo masses}
\label{app1}

This Appendix describes the adopted procedure to match {\pin} and Flagship haloes in AM
and CM (see Sect.~\ref{sec:hod}). Binning haloes in redshift with bin size $\delta
z=0.01$, we measure the cumulative halo mass function of the Flagship simulation (over its
octant footprint) and the stacked one from all the \pin mocks of a set. AM of halo masses
is then obtained by computing the Flagship and {\pin} halo masses that give the same halo
number densities. Because we have only one Flagship simulation, a direct match in number
density would propagate its noise, so we fit the relation between the logarithms of the
two masses with a second-order polynomial; this was found to be very accurate in the halo
mass range of interest, namely from $10^{11}$\,\hmsun to $10^{13}$\,\hmsun. We then
linearly interpolate in redshift the three fitting coefficients of the power laws, thus
creating an AM model for mapping {\pin} halo masses to Flagship halo masses, $M_{\rm AM}$
\be
M_{\rm AM,12} = (a_1 + a_2 z) + (b_1 + b_2 z)\, M_{\rm pin,12} + 
(c_1 + c_2 z)\, M_{\rm pin,12} ^2\, ,
\label{eq:pin2FSmasses}\ee
\noindent
where $M_{\rm AM,12} := \logten (M_{\rm AM}/10^{12}\, \hmsun)$ and $M_{\rm pin,12} :=
\logten (M_{\rm pin}/10^{12}\, \hmsun)$ . The coefficients are given in
Table~\ref{table:pin2FSmasses} for V4 and V5 of the code.

\begin{figure} 
\centering
\includegraphics[width=0.9\hsize]{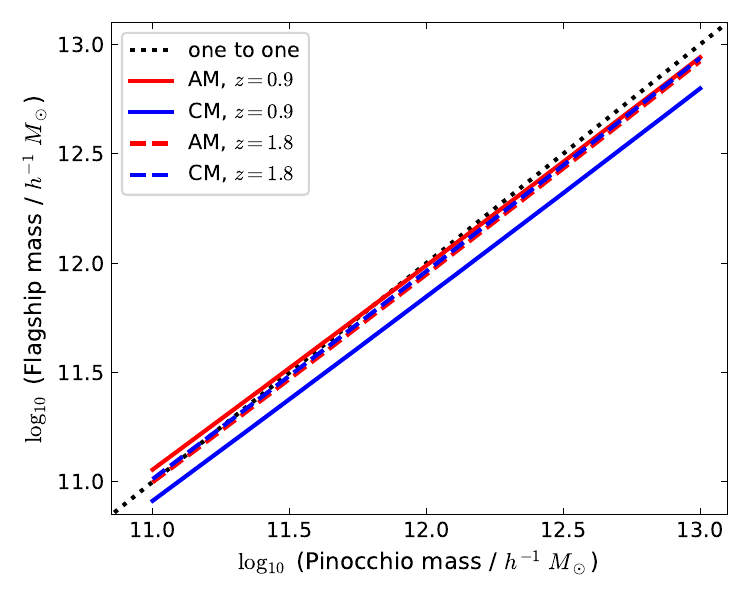}
\caption{Relation between Flagship and {\pin} masses (valid for V5 of the code) in the
  case of AM and CM, at the two redshifts $z=0.9$ and $z=1.8$.}
\label{fig:mass_scaling}
\end{figure}

A brute-force determination of CM would imply to measure power spectra for hundreds of
mock catalogues for many halo mass thresholds and redshifts; we resort to a more
convenient procedure by assuming that the relation between AM and CM masses is well
represented by a redshift-dependent multiplicative shift. This leaves us with two
parameters, that are calibrated using the galaxy catalogues. We require that, at
$k<0.2$\,\kmpc, the average clustering amplitude of 100 {\pin} galaxy catalogues with
$\fha>f_0$, measured in the four redshift bins with edges in $[0.9,1.1,1.3,1.5,1.8]$, is
consistent with the Flagship one projected onto the same {\tdeg} radius footprint, to
within the predicted variance. On smaller scales, that are hardly used by the standard
inference methods but are necessary to constrain the one-halo term, we require that the
clustering amplitude is always reproduced to better than 10\%. After a trial-and-error
procedure based on 100 catalogues, we find that the CM masses of the Flagship mock is well
reproduced with the present scaling
\be \logten (M_{\rm CM}/M_\odot) = \logten (M_{\rm AM}/M_\odot) - 0.125 - 0.175\, (1-z)\, ,
\label{eq:clusteringmatch} \ee
\noindent
with no dependence on the code version. Figure~\ref{fig:mass_scaling} shows the resulting
relation between \pin and Flagship halo masses valid for V5 of the code, in the case of AM
and CM, at redshifts $z=0.9$ and $z=1.8$.

\begin{figure} 
\centering
\includegraphics[width=1\hsize]{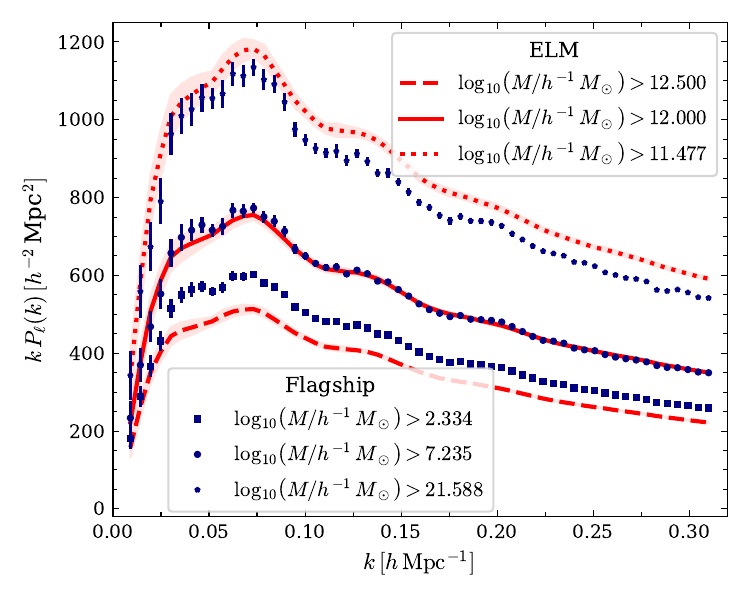}
\caption{Power spectrum monopole of halos, with $z\in [0.9,1.1]$, 
  in the Flagship mock and in 50 
  EuclidLargeMocks for three distinct mass cuts that obey the CM relation.}
\label{fig:halo_clustering}
\end{figure}

This CM guarantees that halos containing the bulk of {\ha} galaxies, namely those with
$M_{\rm h}\sim10^{12}$ \hmsun, have the same clustering level once the matching masses are
used to define the mass cuts. However, as discussed in \cite{Munari2017}, halo bias is not
recovered with sub-percent accuracy because halo construction is not perfect, and this
weakens the clustering signal making its amplitude more similar to that of the matter
field. This effect is stronger for smaller halos, while massive halos are reconstructed
with higher accuracy. Figure~\ref{fig:halo_clustering} shows the halo power spectrum
monopole in the first redshift bin for three mass cuts, where {\pin} and Flagship mass
cuts obey the CM relation. While halo clustering is very well recovered at
$\sim$\,$10^{12}$ \hmsun, the correction is not strong enough for smaller halos while it
is too strong at larger masses. This is in line with the trend shown in
Fig.~\ref{fig:PKflux}. A more sophisticated clustering matching procedure would easily
correct for this difference, but this mismatch has a minor impact on the mock catalogues.

\begin{table}
\caption{Coefficients of the relations that map {\pin} halo masses to Flagship AM masses
  (Eq.~\ref{eq:pin2FSmasses}), for the two code versions used in this paper.}
\begin{center}
\begin{tabular}{l|S|S}
& V4 & V5 \\ \hline
$a_1$ &  0.0243  &  0.0302\\
$a_2$ & -0.0463  & -0.0458\\
$b_1$ &  0.923   &  0.923\\ 
$b_2$ &  0.0220  &  0.0219\\
$c_1$ &  0.00729 &  0.00673\\
$c_2$ &  0.00298 &  0.00299\\
\end{tabular}
\end{center}
\label{table:pin2FSmasses}
\end{table}

\end{appendix}

\label{LastPage}
\end{document}